\documentclass[a4paper,11pt]{article}
\usepackage{jcappub} % for details on the use of the package, please
\usepackage{subfigure} 
\title{\boldmath Effects of neutrino mass and asymmetry on cosmological structure formation}

\author[a, b, c, 1]{Zhichao Zeng,\note{Corresponding author.}}
\author[a]{Shek Yeung}
\author[a]{and Ming-chung Chu}

\affiliation[a]{Department of Physics, the Chinese University of Hong Kong, Sha Tin, NT, Hong Kong}

\affiliation[b]{Department of Physics, the Ohio State University, Columbus, Ohio 43210, USA}

\affiliation[c]{Center for Cosmology and AstroParticle Physics (CCAPP), the Ohio State University, Columbus, Ohio 43210, USA}

\emailAdd{zeng.408@buckeyemail.osu.edu}
\emailAdd{syeung@phy.cuhk.edu.hk}
\emailAdd{mcchu@phy.cuhk.edu.hk}

\abstract{Light but massive cosmological neutrinos do not cluster significantly on small scales, due to their high thermal velocities. With finite masses, cosmological neutrinos become part of the total matter
field and contribute to its smoothing. Structure formation in the presence of massive neutrinos is therefore impeded compared to that in the standard $\Lambda$CDM cosmology with massless neutrinos. Neutrinos' masses also distort the anisotropy power spectrum of cosmic microwave background (CMB). Furthermore, a finite chemical potential $\mu$ for cosmological neutrinos, still allowed by current data, would have a non-negligible impact on CMB and structure formation. We consistently evaluate effects of neutrino masses and chemical potentials on the matter power spectrum by use of a neutrino-involved N-body simulation, with cosmological parameters obtained from a Markov-Chain Moonte-Carlo (MCMC) refitting of CMB data. Our results show that while a finite averaged neutrino mass $m_\nu$ tends to suppress the matter power spectrum in a range of wave numbers, the neutrino degeneracy parameters ${\xi_i \equiv \mu_i /T}$  ($i=$1, 2, 3) enhance the latter, leading to a large parameter degeneracy between $m_\nu$ and $\xi_i$. We provide an empirical formula for the effects on the matter power spectrum in a selected range of wave numbers induced by $m_\nu$ and $\eta \equiv \sqrt{\sum_i \xi^2_i}$. Observing a strong correlation between $m_\nu$ and $\eta$, we propose a single redshift-independent parameter $m_\nu - \frac{4}{3}\eta^2$ to characterize the neutrino effects on the matter power spectrum.}

\begin{document}

\maketitle
\flushbottom

\section{Introduction}
\label{sec:intro}

Cosmological neutrinos are believed to be the most abundant fermions in the Universe. However, this cosmological neutrino background (CNB) is difficult to detect directly, due to its low temperature of about 2 K now. On the other hand, results of neutrino oscillation experiments show that there are three neutrino mass eigenstates, $m_{\nu_i} ( i =$ 1, 2, 3),  with mass-squared splittings $\Delta m^2_{12} \approx 7.37 \times 10^{-5} $ eV$^2$ and $|\Delta m^2_{23}| \approx 2.50 \times 10^{-3} $ eV$^2$ \cite{nu_osc}. These slightly massive neutrinos effectively behave as hot dark matter, which has been ruled out as the dominant part of the dark matter \cite{hdm}, although it is still the only known component. This CNB is expected to have impact on the CMB anisotropies and large-scale structure (LSS) formation. Compared to those in the standard $\Lambda$CDM cosmology with massless neutrinos, the radiation-matter equality time and the Hubble expansion rate would be modified with massive cosmological neutrinos. The photon diffusion scale $\theta_d$ and sound horizon scale $\theta_s$ would then be altered, resulting in a modified angular anisotropy power spectrum of the CMB. The integrated Sachs-Wolfe effect and the gravitational lensing of CMB polarization are also affected by massive neutrinos via early structure growth \cite{aba_nu}. Altogether the updated constraints on the sum of neutrino masses are $\sum_i m_{\nu_i} < 0.23$ eV from the Planck 2015 observation \cite{planck2015}, and $\sum_i m_{\nu_i} < 0.12$ eV from the newest Planck 2018 result \cite{planck2018}. 

The large scale structure formation is more sensitive than CMB to the sum of neutrino masses. It is well known that the growth of structure is governed by the competing effects of the cosmic expansion and self-gravity of matter perturbations, both of which are affected by the massive neutrinos \cite{ab2013}. Unlike photons, slightly massive cosmological neutrinos experience a transition from being relativistic to non-relativistic as the universe cools down, resulting in a different background expansion rate from the $\Lambda$CDM model. On the other hand, compared to cold dark matter (CDM), cosmological neutrinos have a higher thermal velocity and are less likely to be trapped by potential wells. Thus their spatial distribution is much more disperse than CDM. As a consequence, below its free-streaming scale, the CNB slows down the growth of structures, leading to a suppression of the total matter power spectrum. This free-streaming effect has traditionally been studied in the linear regime \cite{cpma, eh}, and included in numerical Boltzmann codes such as \texttt{CAMB} \cite{camb}. However, as the Lagrangian perturbation theory is limited to the condition that the over-density field $|\delta| \ll 1$, this linear method cannot be used to explore the influence of massive neutrinos on the late non-linear growth of structures. Instead, cosmological N-body simulation or simulation-based Halofit formula should be used \cite{halofit}. A natural way to implement neutrinos in N-body simulations is to treat them as an independent kind of simulation particles, with a much higher typical velocity compared to that of CDM particles \cite{brandbyge08, viel10, vnavarro12, tiannu}. This particle-based simulation is in principle accurate but computationally expensive, and the consequent power spectrum is dominated by shot noise in small scales due to the finite number of neutrino particles \cite{ab2013, brandbyge08}. Another much cheaper alternative is to implement neutrinos as an over-density field on regular grids, and its evolution is then studied by linear perturbation theory. This method is justifiable since neutrinos do not significantly cluster below its free-streaming scale, $k_{fs}$, which is larger than the non-linear scale $k_{nl}$ \cite{ab2013}. This grid-based simulation was first proposed in \cite{brandbyge09}, in which the neutrino power spectrum was evolved by \texttt{CAMB}  and improved to include the non-linear effect of CDM clustering in \cite{ab2013}. The consistency between these two approaches to include neutrinos in cosmological N-body simulation has been well tested in both \cite{ab2013} and this paper.

Most of the studies mentioned above use the same cosmological parameter set when comparing the matter power spectra for different neutrino masses. However, the cosmological parameters obtained from fitting the CMB power spectrum also depend on the neutrino mass given. A fully consistent study should take this into consideration. This is regarded as a third mechanism for massive neutrinos to affect the LSS in our work, although it is not really independent from either the expansion history or the free-streaming effect. Because both CMB and LSS are sensitive to the sum of neutrino masses instead of the mass hierarchy \cite{hierarchy}, we add a new variable $m_{\nu} \equiv \sum m_{\nu_i} / 3$, the averaged mass for the three neutrino mass eigenstates, into a Markov-Chain Monte-Carlo code \texttt{CosmoMC} for fittings of cosmological parameters from the Planck 2015 CMB data (as the likelihood code of Planck 2018 is not released to public yet) \cite{planck2015, cosmomc}. N-body simulations are then generated using these sets of refitted cosmological parameters. In this way the study of neutrinos' influence on LSS is self-consistent.

Another mystery about neutrinos is whether they are Majorana or Dirac particles. If we assign the chemical potentials of neutrinos to be $\{\mu_i\}$, where $i$ labels the neutrino mass eigenstates, then for anti-neutrinos they are $\{-\mu_i\}$. Naturally Majorana neutrinos must have $\mu_i\ = 0$, and if $\mu_i\ \neq 0$, neutrinos are Dirac fermions. Because the neutrino distribution has been frozen after decoupling, $\{\xi_i \equiv \mu_i / T\}$ are fixed and denoted as the neutrino degeneracy parameters. The difference between $\{\xi_i\}$ and $\{-\xi_i\}$ leads to an asymmetry in the neutrino and anti-neutrino number densities. Big Bang nucleosynthesis (BBN) constrains this asymmetry of electron-type neutrinos to be small, with $|\xi_e| \leq \mathcal{O} (10^{-2})$, while the total neutrino asymmetry is mainly constrained by the effective number of relativistic species $N_{eff}$ by CMB, and $\xi_{\mu, \tau}$ of $\mathcal{O}(1)$ is still allowed \cite{bbn, barenboim_asym}. In this paper we also include the possibility of finite $\{\xi_i\}$ in our \texttt{CosmoMC} fitting for cosmological parameters as well as the modified N-body simulation to study its influence on LSS, in addition to the effect of neutrino mass. We follow \cite{barenboim_mass_eigenstate} to set $\xi_e = 0$ and $\xi_\mu = \xi_\tau$, the latter because of the strong mixing between $\nu_\mu$ and $\nu_\tau$. Only one free parameter is left for $\{\xi_i\}$, which we choose to be $\eta \equiv \sqrt{\sum \xi_i^2}$.

In this work, we examine the influence of the averaged neutrino mass $m_\nu$ and the neutrino degeneracy parameter $\eta$ on the total matter power spectrum. All three mechanisms: the modification of the cosmic expansion rate, the neutrino free-streaming effect and the shifts in cosmological parameters obtained from CMB fitting are consistently included. We mainly use the grid-based method of including neutrinos in N-body simulation, by our own modified version of the \texttt{Gadget2} code \cite{gadget2}. 

This paper is organized as follows. In Section 2 we elaborate on the calculation of neutrino energy density and the treatment of $\{\xi_i\}$ with known constraints. The introduction to our \texttt{CosmoMC} refitting is also included. We elaborate on the detailed procedure to conduct both the particle-based and grid-based simulations, and compare the measured total matter power spectra to show the consistency of these two methods in Section 3. We present our results in Section 4, and provide an empirical formula for the neutrino induced change in the matter power spectrum. Our summary and disscussion are in Section 5. 

\section{Neutrino energy density and the cosmic expansion}
\subsection{Cosmic neutrino background}

Cosmological neutrinos are thermally produced relic particles of the Big Bang, and they follow the Fermi-Dirac distribution:

\begin{equation}
f_{\nu} (E, T) = \frac{1}{e^\frac{E - {\mu}}{T}+1},
\end{equation}
where $E = \sqrt{p^2+m^2}$ is the neutrino energy.

The neutrino energy density is given by 

\begin{equation}
\rho_{\nu}(T) = \frac{1}{2\pi^2 \hbar ^3} (\int _0 ^\infty \frac{E}{e^\frac{E - {\mu}}{T}+1} p^2 dp
 + \int _0 ^\infty \frac{E}{e^\frac{E + {\mu}}{T}+1} p^2 dp).
\end{equation}
 
The normal number of degrees of freedom $g_{\nu} = 2$ is separated into the neutrino and anti-neutrino terms here. Cosmic neutrinos decouple from baryons at temperature $T \sim 1$ MeV, and the thermal distribution has been frozen ever since, i.e. the denominator $e^{\frac{E-\mu}{T}} +1$ is fixed. Then the chemical potential $\mu$ scales as $T$, and so we have a constant degeneracy parameter $\xi \equiv \mu / T$.  Because at $T \sim$1 MeV neutrinos are highly relativistic, $E$ is simply replaced by the momentum $p$ in the distribution fuction (2.1). Finally we have:

\begin{equation}
\rho_{\nu}(T) = \frac{1}{2\pi^2 \hbar ^3}(\int _0 ^\infty \frac{E}{e^{\frac{p}{T}-\xi}+1} p^2 dp
 + \int _0 ^\infty \frac{E}{e^{\frac{p}{T} + \xi}+1} p^2 dp).
\end{equation}

The neutrino temparature $T_\nu$ is well known to be related to the CMB temparature $T_\gamma$
\begin{equation}
T_{\nu} = (\frac{4}{11}) ^{1/3} T_{\gamma}. 
\end{equation}

This relation is derived from the entropy conservation before and after the electron-positron annihilation, which heats up the photons but not the neutrinos. Therefore $T_{\nu}$ should not be directly dependent on $m$ or $\xi$. On the other hand, the neutrino degeneracy $\xi$ reduces the weak interaction rate between neutrinos and other species, because some of the initial and final states are occupied \cite{lesg}. Therefore, the neutrino decoupling temperature $T_{dcp}(\xi)$ would be higher. But since the derivation of (2.4) follows the simplified model that neutrinos had completely decoupled from other species before the $e^- e^+$ annihilation and were not heated, which still holds for finite $\xi$, so Eq. (2.4) is still valid.

\subsection{Neutrino asymmetry and constraints on $\{\xi_i\}$}

A finite $\xi$ would naturally lead to an asymmetry of neutrinos and anti-neutrinos, defined by

\begin{equation}
L \equiv \frac{n-\bar{n}}{n_{\gamma}} \propto [\xi ^3 + \pi^2 \xi],
\end{equation}

where $n$, $\bar{n}$ and $n_\gamma$ are the number densities for neutrinos, anti-neutrinos and photons respectively \cite{lesg}. Eq. (2.5) is exact for massless neutrinos, and it is still good approximation for light but massive neutrinos, with an error of $\mathcal{O}(1\%)$ for $m < 0.1$ eV. The asymmetry of electron neutrinos $L_e$ would affect the neutron-proton ratio, which is tightly bounded by the ${}^{2}$H/${}^{1}$H ratio and the ${}^{4}$He abundance in BBN \cite{bbn2010}. The updated constraint gives $-0.018 \leq \xi_e \leq 0.008$ ($-4.5\leq 10^3 L_e \leq 2.0$) at 68\% C.L. On the other hand, $\xi_{\mu}$ and $\xi_{\tau}$ of $\mathcal{O}(1)$ are still allowed, being only weakly constrained by the extra number of relativistic species $\Delta N_{eff} = N_{eff} - 3.046$ by both CMB and BBN \cite{planck2015, bbn, barenboim_asym, barenboim_mass_eigenstate}. 

In the early universe, the high interaction rate blocks neutrino flavor oscillations and keeps neutrinos in flavor eigenstates. Thus the asymmetry matrix $L$ is diagonal in $L_{\alpha}$ ($\alpha = e, \mu, \tau$). However, as shown in \cite{barenboim_mass_eigenstate}, as temperature drops below $\sim 15$ MeV, neutrino flavor oscillations become active and off-diagnoal components in $L_{\alpha}$ become significant. At around $T \sim 2$ -- 5 MeV, right before BBN and neutrino decoupling, the evolution of $L_{\alpha \beta}$ reaches equilibrium again, and $L$ is diagonal in mass eigenstates, $L_i$ ($i=1,2,3$). We can use the Pontecorvo-Maki-Nakagawa-Sakata (PMNS) matrix to transform between the flavor and mass bases:

\begin{equation}
L_i = U_{PMNS}^{-1} L_\alpha U_{PMNS}.
\end{equation}

More specifically we have 

\begin{equation}
L_e  = c_{13}^2(c_{12}^2 L_1 + s_{12}^2 L_2)  + s_{13}^2L_3, 
\end{equation}

where $c_{ij}, s_{ij}, t_{ij}$ are the cosine, sine and tangent of the mixing angle $\theta_{ij}$. Assuming $L_e = 0$, the asymmetries of other flavors become 

\begin{equation}
\begin{split}
L_{\mu} = c_{23}[c_{23}(1-t_{12}^2) &- 2\cos(\delta_{CP})s_{13}s_{23}t_{12} ] L_2 \\
 + \{s_{23}^2(1-t_{13}^2) &- t_{13}^2  t_{12}c_{23} [ t_{12} c_{23} + 2\cos(\delta_{CP}) s_{13}s_{23}) ]\}L_3, \\
L_{\tau} = s_{23} [s_{23}(1-t_{12}^2) & + 2\cos(\delta_{CP}) s_{13}c_{23}t_{12}]L_2  \\ + \{c_{23}^2 (1-t_{13}^2)  + & t_{12} t_{13}^2  s_{23}[2\cos (\delta_{CP}) s_{13}c_{23} - t_{12}s_{23}]\} L_3, \\
\end{split}
\end{equation}
where the neutrino CP phase $\delta_{CP}$ is relevant.

Furthermore, due to the strong mixing of $\nu_\mu$ and $\nu_\tau$, we follow \cite{barenboim_mass_eigenstate} to set $L_\mu = L_\tau$. Thus in the set of equations (2.7) and (2.8) we only have one degree of freedom left to determine $L_i$ and $\xi_i$,  which we choose to be $\eta \equiv \sqrt {\sum \xi_i^2}$. This is because the main CMB constraint on the total asymmetry is the deviation of the effective number of relativistic degrees from 3 (instead of 3.046 because the small distortion of 0.046 caused by incomplete neutrino decoupling before the $e^- e^+$ annihilation should not be affected much by finite $\{\xi_i\}$)

\begin{equation}
\Delta N_{eff} = (\frac{\sum_i \rho_{\nu_i} + \sum_i \rho_{\bar{\nu_i}}}{\frac{7}{8} ( \frac{T_\nu}{T_\gamma})^{4} \rho_{\gamma} } -3) \propto \sum_i(\xi_i ^2 + \frac{\xi_i^4}{2\pi^2}),
\end{equation}

which is proportional to $\eta^2$ up to the leading order, independent of the CP phase $\delta_{CP}$.

\subsection{Refitting the Planck 2015 data with \texttt{CosmoMC}}

The fiducial model of fitting from CMB includes a cosmological parameter set of six free variables: $\{\Omega_\mathrm{c}h^2,~\Omega_\mathrm{b}h^2,~\theta,~\tau,~A_\mathrm{s},~n_\mathrm{s}\}$.
Besides, it is also a common practice, as in the Planck 2015 paper \cite{planck2015}, to include $\sum m_{\nu_i} (=3m_\nu)$ as another free variable in the MCMC process to set constraints on it. 

In this work, apart from the sum of neutrino masses $\sum m_{\nu_i}$, we modify \texttt{CAMB} to include the calculations of neutrino
energy densities with non-zero $\eta$ \cite{camb}. So we have a set of 8 parameters as our variables of \texttt{CosmoMC}: $\{\Omega_\mathrm{c}h^2,~\Omega_\mathrm{b}h^2,~\theta,~\tau,~A_\mathrm{s},~n_\mathrm{s},~m_\nu,~\eta\}$.
In section (4.1) we use the mean values of the refitting results when $\{m_\nu, \eta\}$ are both freely varying, while in section (4.2) we sample different sets of $\{m_\nu, \eta\}$ with equal spaces and fix them in \texttt{CosmoMC}, so as to study the degeneracy of their effects. 

This modified
\texttt{CosmoMC} is used together with the Planck 2015 data and likelihood codes \texttt{lowTEB} and \texttt{plikHM\_TTTEEE} to find the best-fit and mean values of cosmological parameters \cite{planck2015, cosmomc}. More
details can be found in (Lau et al. in preparation) and Appendix C, where we show the fitting results of cosmological parameters with or without $\eta$ as a free parameter (the fiducial model here includes $m_\nu$). We can also see that the tension of $H_0$ is alleviated when $\eta$ is included in the MCMC fitting.
 
\subsection{Cosmic neutrinos and background expansion} 
 
Eq. (2.3) clearly reduces to $\rho \propto T^4 \propto a^{-4}$ and $\rho \propto a^{-3}$ at the ultra-relativistic ($p \gg m$) and non-relativistic ($p \ll m$) limits respectively. We define a power law index $x$ such that $\rho_{\nu} \propto a^{x}$, with $\eta$ and $m_\nu$ to be parameters, and we show $x$ as a function of $z$ in Fig. 1a, with parameters listed in Table 1.
 
\begin{table}[tbp]
\centering
\begin{tabular}{ |c c | c | c |c |c  c  c |}
\hline
no. & model & $m_{\nu}$(eV) & $\eta$ & $H_0$ & $\Omega_{c+b}$ & $\Omega_{\nu}$ & $\Omega_{\Lambda}$\\
\hline
A1 & fiducial & 0 & 0 & 67.74 & 0.3097 & $\sim 10^{-5}$ & 0.6844\\
A2 & $\eta$ &0 & 0.359 & 67.74 & 0.3097 & $\sim 10^{-5}$ & 0.6844\\
A3 & $m_\nu$ & 0.048 & 0 & 67.74 & 0.3064 & 0.0033 & 0.6844\\
A4 & $m_\nu, \eta$ & 0.048 & 0.359 & 67.74 & 0.3063 & 0.0034 & 0.6844\\
A5 & $2m_\nu$ &0.096 & 0 & 67.74 & 0.3030 & 0.0067 & 0.6844\\
A6 & $2m_\nu, \eta$ &0.096 & 0.359 & 67.74 & 0.3029 & 0.0068 & 0.6844\\
\hline
A7 & $m_\nu, \eta$, refitting & 0.048  & 0.359 & 66.95 & 0.3203 & 0.0035 & 0.6762\\
\hline
\end{tabular}
\caption{\label{table 1} Parameters for comparison of the modified Hubble parameter $H(m_{\nu}, \eta)$ and that in the CDM cosmology $H(0,0)$. Data A1 is the pure CDM case. For A2-A6, we sampled $m_{\nu}$ to be 0, $0.048$ and $0.096$ eV, and $\eta = 0.0,\ 0.359$. We follow the convention to keep $\Omega_m = \Omega_{c+b} + \Omega_{\nu}$ a fixed value, and so the excess of neutrino energy density is effectively deducted from the fraction of dark matter. In A7, all the cosmological parameters are consistently refitted from the Planck 2015 data using our \texttt{CosmoMC} code, with $m_{\nu}$ and $\eta$ included.}
\end{table} 
 
We sample $m_\nu$ with values 0, 0.048 eV and 0.096 eV, which are 0, 1, 2 times the mean value of $m_\nu$ in  our \texttt{CosmoMC} fit of the Planck 2015 data. We also run for $\eta = 0, 0.359$, the latter being the mean value in the \texttt{CosmoMC} fit. As shown in Fig. 1a, the cosmological neutrinos clearly experience a smooth transition from $x=-4$ to $x=-3$, roughly from redshift $z \sim 10^4$ to $z \sim 10^1$, the cosmic redshifts most important for cosmological structure formation. The transition occurs earlier for more massive neutrinos, as expected. It can also be seen from Fig. 1a that $\eta$ does not have much effect on the transition. 

There is naturally a deviation of the Hubble expansion history from that of the standard $\Lambda$CDM Universe. The Hubble expansion rate is determined by the Friedmann equation

\begin{equation}
H \equiv \frac{\dot{a}}{a} = \sqrt{ \frac{8\pi G}{3} (\rho _{\gamma} + \rho _{\nu} + \rho _{c+b} + \rho _{\Lambda})},
\end{equation}

where $\rho_{\Lambda} \propto a^0$, $\rho_{c+b} \propto a^{-3}$ and $\rho_{\gamma} \propto a^{-4}$ are the energy densities of dark energy, matter, and radiation respectively. We have to evaluate the integration (2.3) numerically for $\rho_\nu$, due to the transition. The resulting ratio $H(m_{\nu}, \eta)/H(0, 0)$ versus $z$ is plotted in Fig. 1b for the parameter sets in Table 1.

\begin{figure}
\subfigure[Fig. 1a]{
\includegraphics[width=8cm, height=7cm]{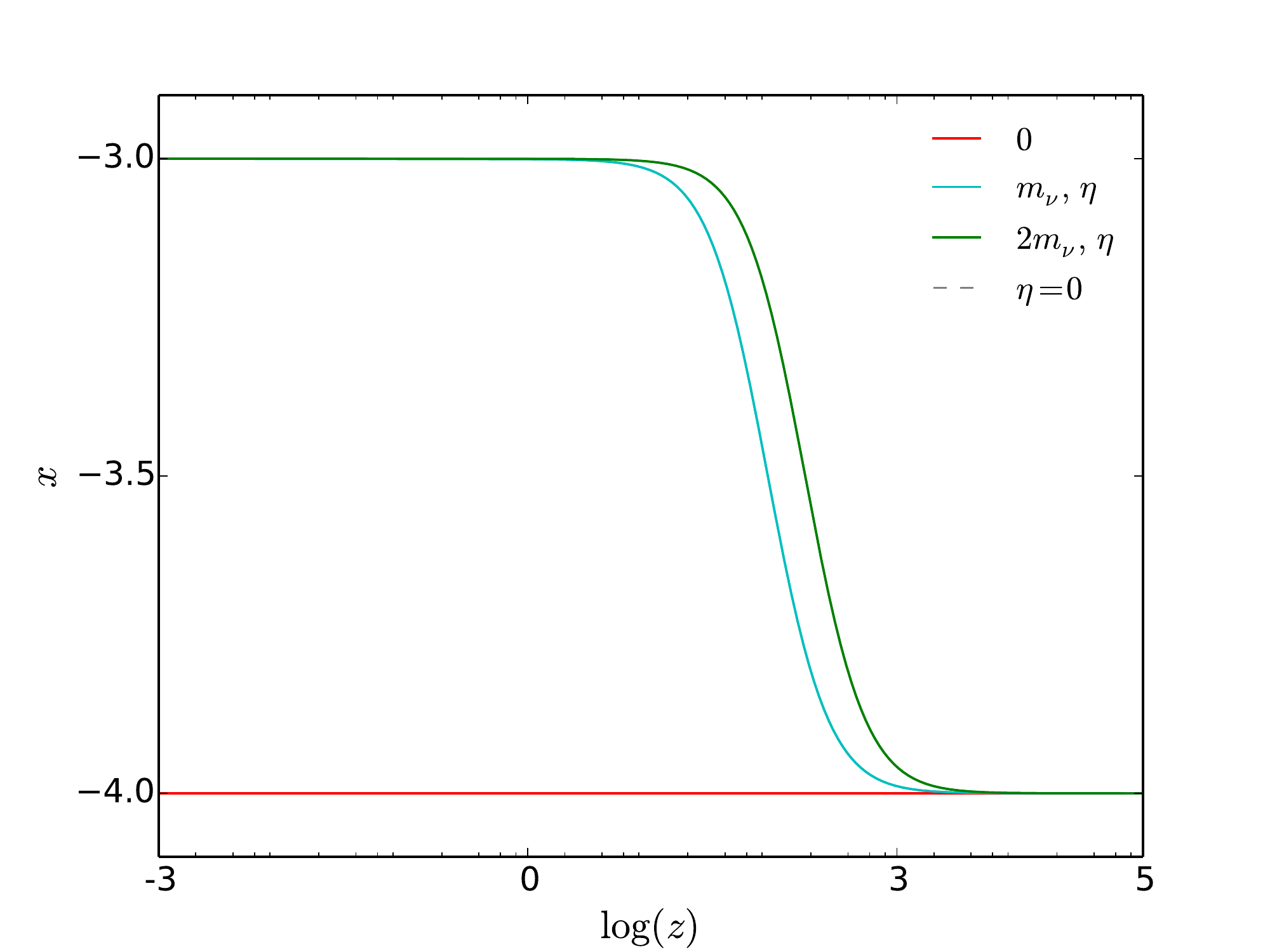}
}
\subfigure[Fig. 1b]{
\includegraphics[width=8cm, height=7cm]{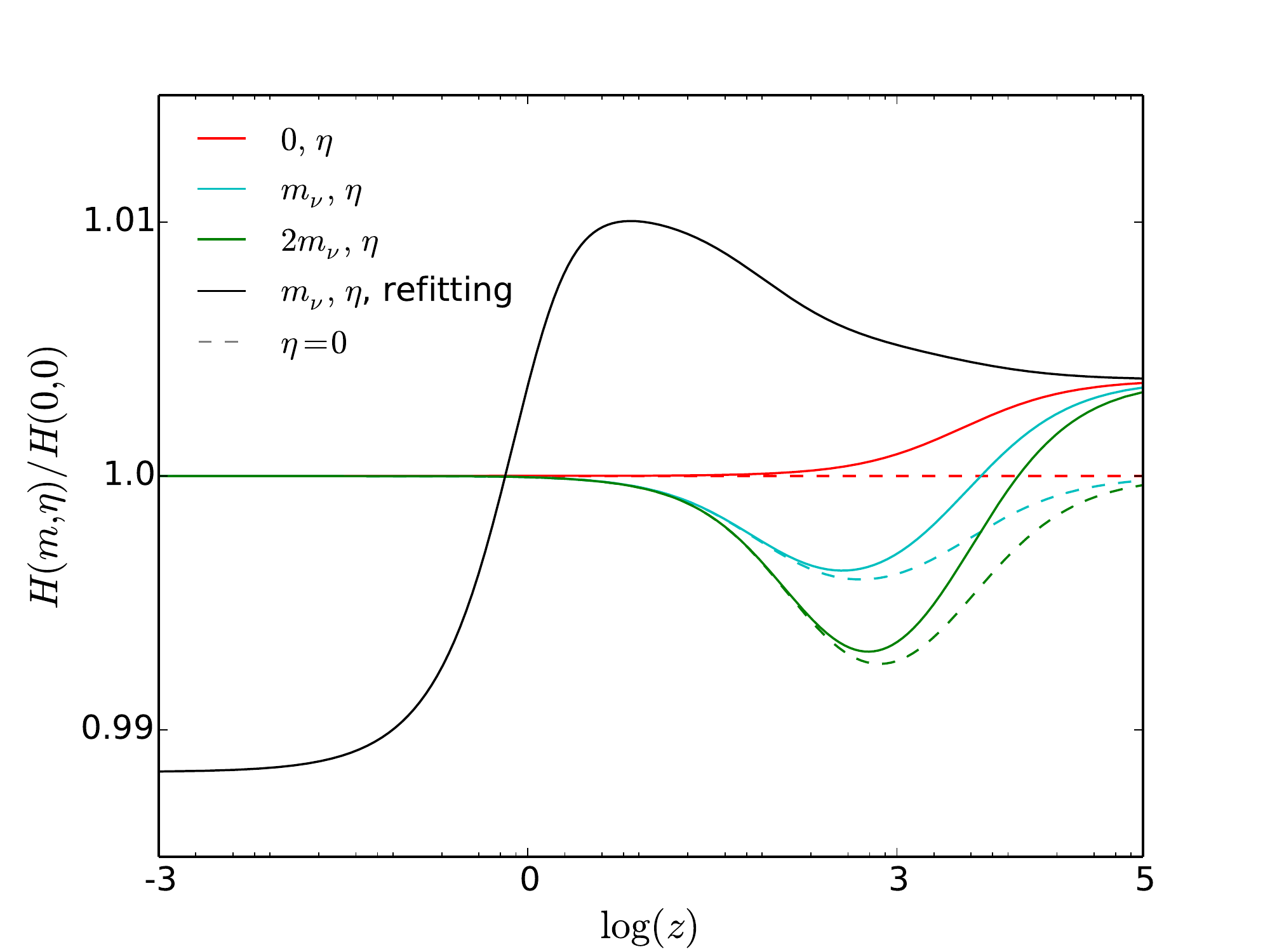}
}
\subfigure[Fig. 1c]{
\includegraphics[width=8cm, height=7cm]{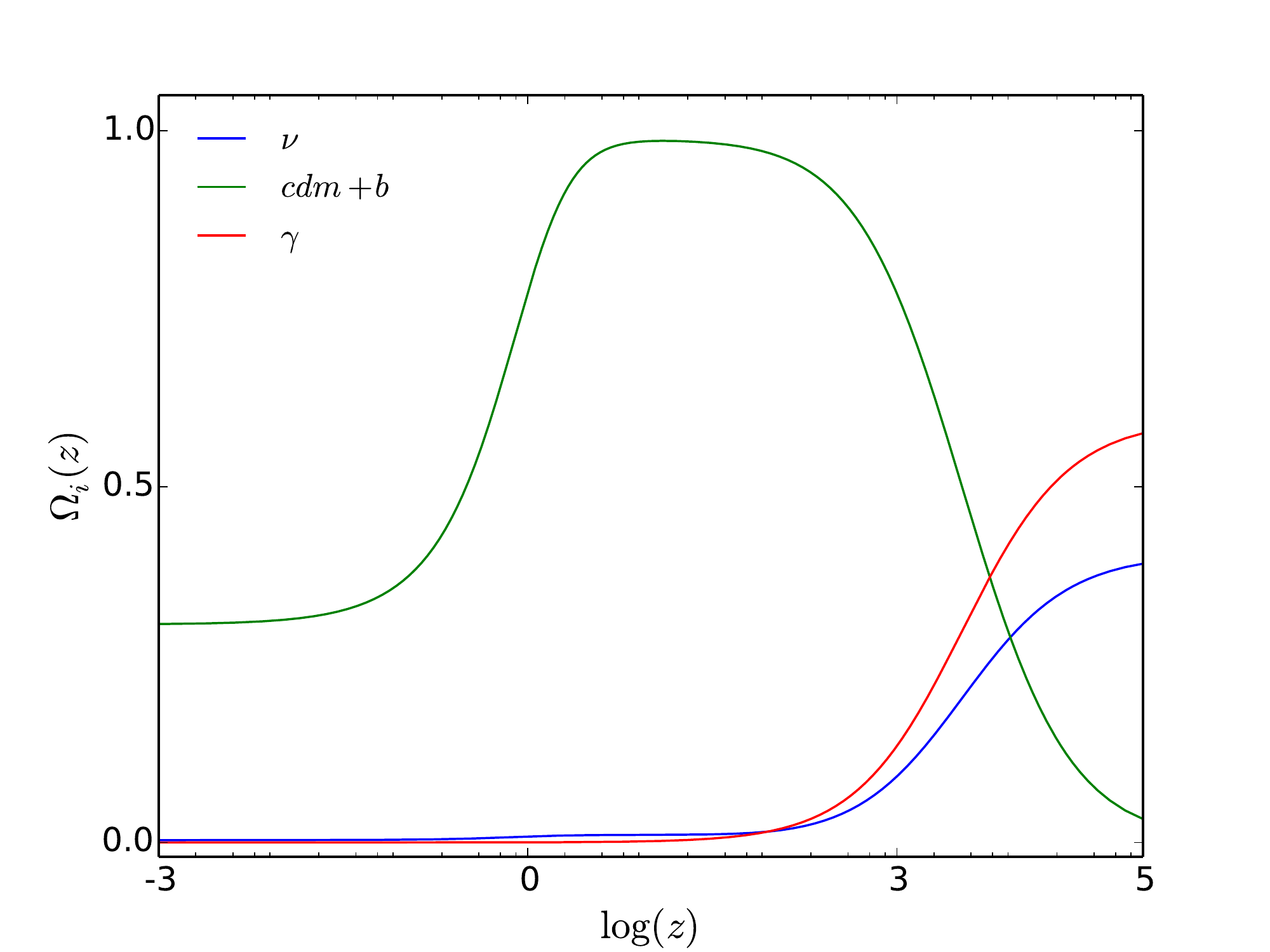}
}
\subfigure[Fig. 1d]{
\includegraphics[width=8cm, height=7cm]{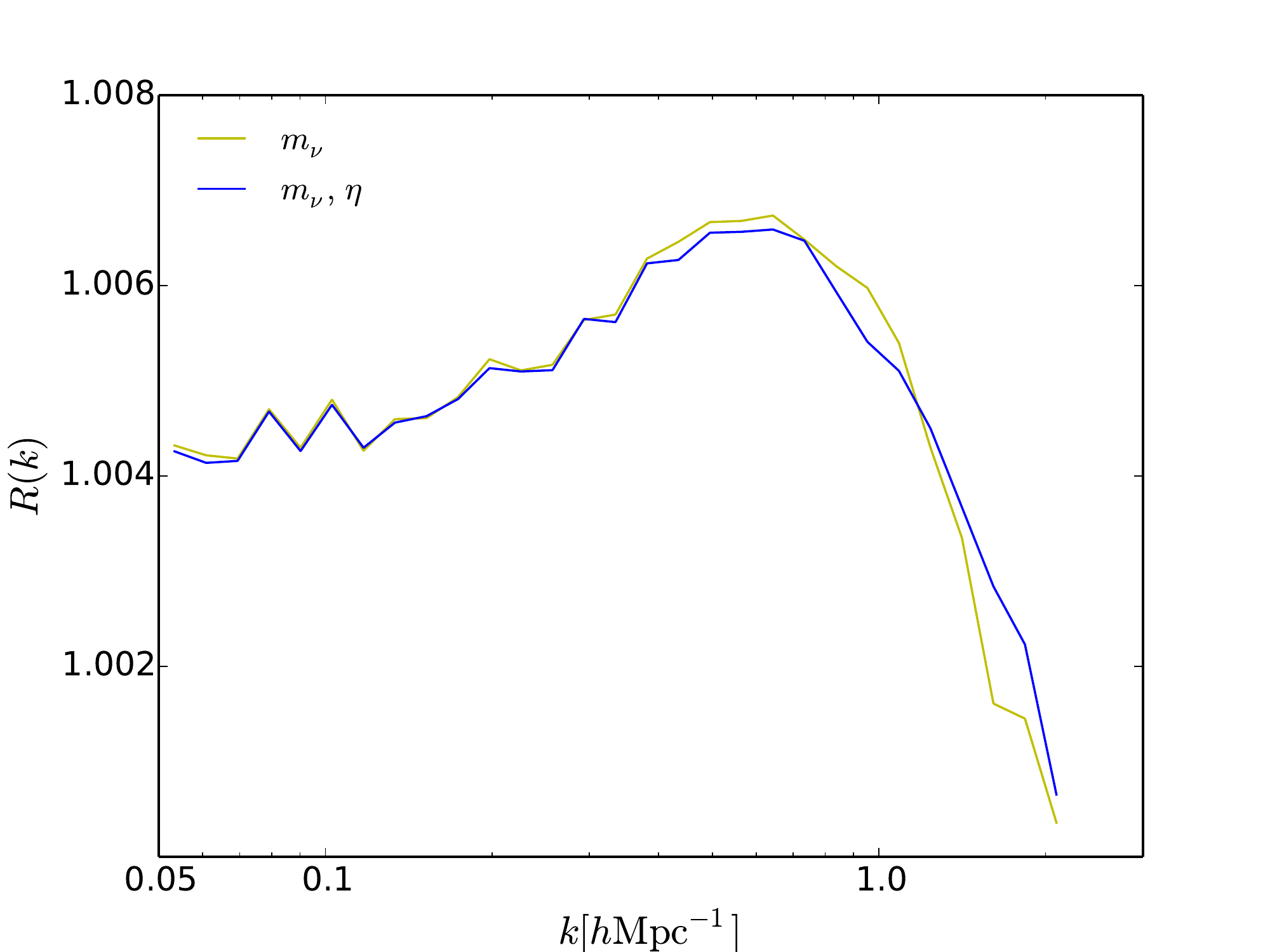}
}
\caption{\label{fig1} a) Relation between the power index $x$ of neutrino energy density and redshift $z$. The dashed red, cyan and green curves show $x$ for data sets A1, A3 and A5, with $m_{\nu} = 0, 0.048, 0.096$ eV respectively and $\eta = 0$. The corresponding cases for $\eta = 0.359$ are shown in solid curves, which are indistinguishable from the dashed ones. b) Ratio between the modified Hubble expansion rate $H(m_\nu, \eta)$ and that of $\Lambda$CDM $H(0,0)$, for parameter sets A1 (red dashed), A3 (cyan dashed), A5 (green dashed), A2 (solid red), A4 (solid cyan) and A6 (solid green). The solid black curve is for A7, in which $m_{\nu} = 0.048$ eV, $\eta$ = 0.359, and the cosmological parameters are refitted from Planck 2015 data using \texttt{CosmoMC}. c) Energy density fractions $\Omega_i$ of matter (green), photons (red) and neutrinos (blue) respectively, for $m_\nu = 0.048$ eV and $\eta =0$. d) Relative matter power spectra with influence of neutrinos, when only the modified Hubble expansion rate is considered. The yellow curve shows the power spectrum of A3, with $m_{\nu} = 0.048$ eV, relative to the power spectrum of A1, which is from the standard CDM cosmology. The blue curve shows the ratio of the power spectra for A4 and A1.}
\end{figure}

For parameter sets A1-A6, the same cosmological parameters are used, with $\Omega_i$ and $H_0$ fitted from our \texttt{CosmoMC}, fixing $(m_{\nu}, \eta) = (0, 0)$. When neutrinos are massive and $\Omega_{\nu}$ not negligible, we fix $\Omega_m = \Omega_{c+b} + \Omega_{\nu}$ as a constant, so as to keep a flat Universe. It is the conventional treatment in previous studies to deduct the excess due to neutrino energy density from cold dark matter \cite{ab2013, halofit, brandbyge09}. 

In Fig. 1b we can see that when $\eta > 0$ with zero neutrino mass, $H$ is increased at high redshift, because the effective number of relativistic species $N_{eff}$ is larger. Corresponding cases for $H(m_{\nu}, 0)$ are all lower than $H(0,0)$, as shown by the dashed curves in Fig. 1b. This suppression peaks at $z \sim 10^3$, with a magnitude $-0.4\%$ for $m_{\nu} = 0.048$ eV and is roughly proportional to $m_\nu$. To understand this drop, we plot the fractional energy densities $\Omega_{c+b}(z)$, $\Omega_{\nu}(z)$ and $\Omega_{\gamma}(z)$ in Fig. 1c, in which three cosmic eras can be clearly distinguished: radiation-dominated, matter-dominated and dark energy-dominated. From $z \sim 10^4$, the contribution of matter to the total energy density rapidly increases and surpasses radiation. But at this time massive neutrinos are still radiation-like, just entering the transition to matter-like particles, as shown in Fig. 1a. This time delay between the transition of neutrino relativistic degree and the transition of radiation-dominated to matter-dominated era leads to a decrease in total energy density and thus $H$ compared to those in the $\Lambda$CDM cosmology. Later on, this deficiency in total energy density is alleviated as neutrinos evolve to be more matter-like, and the magnitude of the suppression in $H$ gradually returns to zero.

The black solid curve in Fig. 1b (A7/A1 in Table 1) is based on our refitted cosmological parameter set, with $m_{\nu} = 0.048$ eV and $\eta = 0.359$ as parameters. From the row A7 in Table 1, we can see that $\Omega_i$ and $H_0$ are shifted by a noticeable amount. In this way we do not need to worry about the excess of neutrino energy density, as all the cosmological parameters are consistently refitted. The resulting $H(m_{\nu}, \eta)$ behaves distinctly from previous comparisons. Naturally, the normalization factor $H_0$ is different from the $\Lambda$CDM case. Furthermore, the larger $\Omega_m$ after the refitting naturally leads to a peak of $H(m_{\nu}, \eta) / H(0, 0)$ in the matter-dominated era. 

The cosmic structure formation is known to be governed by the competing effects of gravity and cosmic expansion. So the modified $H$ would naturally affect this process. We implement $H(m_{\nu}, \eta)$ in the N-body simulation code \texttt{Gadget2} to study this effect, and the resulting relative matter power spectra $R(k) = P(m_\nu, \eta, k) / P(0, 0, k)$ are shown in Fig. 1d. For A3 and A4 sets of parameters, a slightly stronger clustering of simulation particles can be observed. This is due to the small decrease in expansion rate from $z=49$, when our simulation is initiated. As the power spectrum is affected only at sub-percent level when only the background expansion is modified, we do not separate this effect from neutrino free-streaming in the following discussions.

\section{Neutrino free-streaming effect}

The high thermal speed of cosmological neutrinos makes them weakly bounded to gravitational potential wells provided by matter. So the over-density field of massive neutrinos $\delta_{\nu}$ grows much slower than that of CDM and baryons $\delta_{c+b}$. This leads to a smaller total over-density field $\delta_{t}$ due to the averaging effect:

\begin{equation}
\delta_t = (1-f_{\nu}) \delta_{c+b} + f_{\nu}\delta_{\nu}, 
\end{equation}

where $f_{\nu} \equiv \Omega_{\nu} / \Omega_m$ is the mass fraction of cosmic neutrinos. However, as both neutrinos and matter particles are sources of gravitational potential, one cannot assert that the final over-density field is affected only by $\sim f_{\nu}$. Non-linear effects can amplify the role of massive neutrinos in structure formation.

In the linear regime where $\delta_t \ll 1$, $\delta_{c+b}$ and $\delta_{\nu}$ can be co-evolved via perturbation theory by Boltzmann codes such as  \texttt{CAMB}. Nevertheless, for a complete study including the non-linear effect at late time, N-body cosmological simulation should be used. 

We have adopted two methods for including the neutrino effects in an N-body simulation. In the grid-based method, cosmic neutrinos are treated as a density field on grids, and the total over-density field in Eq.(3.1) is calculated at every simulation time step to keep track of the correct gravitational potential. The evolution of $\delta_{\nu}$ is calculated by the linear theory, the detailed derivation of which is based on \cite{xiang} and discussed in Appendix A. The procedures of our grid-based simulation are listed in Section [3.2]. In the particle-based method, neutrinos are added as a new kind of simulation particles, which have a velocity dispersion different from the CDM particles. We elaborate in detail the setup of our particle-based simulation in Section [3.3], and compare it to the grid-based simulation in Section [3.4].

In principle, the particle-based method should be accurate and is still widely used in recent large-size simulations such as \cite{tiannu}, because it keeps track of the positions and velocities of neutrino simulation particles. But it has two intrinsic disadvantages. First, it is much more computationally expensive, due to not only the increased number of particles, but also the fact that smaller time steps are needed to resolve the motion of the high-speed neutrinos. Second, unlike CDM, neutrinos themselves do not significantly cluster at small scales. Thus the neutrino power spectrum at small scales is dominated by shot-noise. Another problem is that the neutrino chemical potential gives rise to a degeneracy pressure, which cannot be trivially transformed to a particle-particle interaction and implemented into the Newtonian force term in an N-body simulation. Therefore, we choose the grid-base method to be our main simulation strategy, with the particle-based method a reference for comparison.

\subsection{Linear evolution of the neutrino over-density}

From \cite{xiang} the linear growth equation of the neutrino over-density field is 

\begin{equation}
\widetilde{\delta_{\nu}}(s, \textbf{k}) = 4\pi G \int _0 ^s a^4(s')(s-s') \Phi [\textbf{k} (s-s')] [\bar{\rho}_{c+b} (s') \widetilde{\delta}_{c+b}(s', \textbf{k}) + \bar{\rho}_{\nu} (s') \widetilde{\delta_{\nu}}(s', \textbf{k})] ds' + \Phi (\textbf{k} s) \widetilde{\delta_{\nu}}(0, \textbf{k}),
\end{equation}

where $s$ is the time in co-moving coordinate defined in Eq.(A.8), $\widetilde{\delta}_i$ is the over-density of component $i$ in $k$-space, and the function $\Phi$ is given by

\begin{equation}
\Phi (\textbf{q}) = \frac{B_0 + \sum _{n=1}^{\infty}(-1)^{n+1} \{ \xi B_1(n) \sin(A\xi) + \xi B_2(n) \cos(A\xi) + B_3(n) \sin(A\xi) + B_4(n)[ \cos(A\xi) + e^{-n\xi}] \} }{A(\int _0 ^\infty \frac{x^2 }{e^{x-\xi} -1} dx + \int _0 ^\infty \frac{x^2 }{e^{x+\xi} +1} dx)}.
\end{equation}

In Eq.(3.3), $A \equiv qT/m$, $x \equiv um/T$, with $T$, $m$ and $u$ the neutrino temperature, mass and speed respectively. $B_i(n)$ are given in Appendix A.

In Eq.(3.2), the second term on the right hand side is the linear part of the growth equation, so that the initial condition can be separated from a time dependent growth factor. The first term is a complicated self-involved integration, which reflects the interaction between the total gravitational potential and neutrino over-density field. Since $\widetilde{\delta _{\nu}} (s, \textbf{k})$ itself appears in the integral, we need to solve this integral equation iteratively. Now if we have the initial condition $\widetilde{\delta}_{\nu} (0, \textbf{k})$ and the over-density of CDM (together with baryons) $\widetilde{\delta}_{c+b} (s', \textbf{k})$ as a function of time $s'$, we can calculate the neutrino over-density at the final time $s$,  $\widetilde{\delta}_{\nu} (s, \textbf{k})$. We elaborate on the derivation of Eqs.(3.2) and (3.3) in Appendix A.

\subsection{Grid-based neutrino simulation}

\texttt{Gadget2} is a hybrid code of the Tree and PM algorithms. Since we only have CDM particles and simulate cosmic neutrinos as grid-based density field, only the PM part, which is responsible for the long-range force, is re-calculated using Eq.(3.1). The over-density field of CDM and baryons $\delta_{c+b}$ is directly measured in the simulation, while $\delta_{\nu}$ is evaluated by Eq.(3.2). The initial condition $\widetilde {\delta}_{\nu}(0, \textbf{k})$ is recorded in the previous time step, while for $\widetilde {\delta}_{c+b}(s', \textbf{k})$ we use a linear interpolation between $\widetilde {\delta}_{c+b}(0, \textbf{k})$ and $\widetilde {\delta}_{c+b}(s, \textbf{k})$.

Normally in cosmological simulations, the number of PM grids is not smaller than the particle number, which is of $\mathcal{O}(10^6)$ in our calculation. It would be very time consuming to ergodically go through all the $\textbf{k}$-grids. So we use following tricks in \cite{ab2013} to reduce the number of calculations. First, in the growth equation (3.2), there is no mixing between the real and imaginary parts of $\widetilde{\delta_{\nu}}(\textbf{k})$, and so only the growth of Re$[\widetilde{\delta_{\nu}}(\textbf{k})]$ is needed. The complex phase of $\widetilde {\delta _\nu} ({\bf k})$ is succeeded from the initial condition, which we assume to be the same as the phase of the CDM density field from the beginning of the simulation. Second, from Eq.(3.3), $\Phi(\textbf{q})$ only depends on $|\textbf{q}|$ as the directional information is averaged out in the integration. So we only evaluate the evolution of the ensemble average $\widetilde{\delta_{\nu}}(k)$, with a proper binning.

The detailed procedure of our grid-based neutrino simulation is listed as below:

1. The original Boltzmann code \texttt{CAMB} is used to generate the power spectrum of CDM or massive neutrinos or their weighted average at any redshift. We modify the calculation of the neutrino energy density in \texttt{CAMB} so that it also includes the contribution of a non-zero neutrino degeneracy parameter $\xi$.

2. We generate the power spectra for CDM (and baryons) and massive neutrinos separately at the starting redshift $z=49$ using \texttt{CAMB}. The ratio of these two power spectra $r_{ini}(k)$ is also recorded. Then the CDM power spectrum $P_{c+b}(k)$ at $z=49$ is put into the initial condition generator \texttt{2LPT}, which is based on the second order Lagrangian perturbation theory, to generate the initial conditions for our cosmological simulation \cite{2lpt}. Please note that since we already generate the initial power spectra for CDM and neutrinos at the starting redsfhit $z=49$, we have not made use of the calculation of linear growth rate in \texttt{2LPT}, which is incorrect because the effect of massive neutrinos is not implemented in the standard 2LPT formalism. In other words, we only use the calculation of the displacement field and corresponding velocity field in \texttt{2LPT}, given an input power spectrum.

3. At the 0th time step, we calculate the CDM power spectrum $P_{c+b}(k)$, with proper binning $\{k_i\}$. Then the neutrino power spectrum $P_{\nu}(k)$ is given using $r_{ini}(k)$. $\widetilde{\delta}_{c+b}(k) = \sqrt{P_{c+b}(k)}$ and  $\widetilde{\delta}_{\nu}(k) = \sqrt{P_{\nu}(k)}$ are saved as the initial conditions for the first calculation of Eq.(3.2).

4. At the $n$th ($n \geq 1$) time step, we have the initial over-density fields for both CDM and neutrinos, $\widetilde{\delta}_{c+b}(0, k)$ and $\widetilde{\delta}_{\nu}(0, k)$, recorded in the previous time step. Then we calculate the CDM over-density at the current time step $\widetilde{\delta}_{c+b}(s, k)$ using \texttt{Gadget2}. Here the time $s$ in co-moving coordinate is calculated from $ds = \frac{dt}{a^2(t)}$, as shown in Eq.(A.8). We have

\begin{equation}
 s = \int _{a_{n-1}} ^{a_{n}}\frac{1}{a^3H(a)}da, 
 \end{equation} 

where $a_{n-1}$ and $a_{n}$ are the scale factors of the previous and current PM time steps. Then $\widetilde{\delta}_{c+b}(s', k)$ at any time $s'$ in $[0, s]$ is obtained by linear interpolation, since the time difference between two PM time steps is usually small.

5. Then we put $\widetilde{\delta}_{c+b}(s', k)$ and $\widetilde{\delta}_{\nu}(0, k)$ into the Volterra equation Eq.(3.2), and solve it iteratively with $\Phi(ks) \widetilde{\delta}_{\nu}(0, k)$ as the initial trial function. Usually one iteration would be enough for convergence. Finally we have the neutrino over-density at the current time step $\widetilde{\delta}_{\nu}(s, k)$.

6. Following the assumption that the over-density fields for CDM and neutrinos have the same phase, we can restore $\widetilde{\delta}_{\nu}$ on each $\textbf{k}$-grid:

\begin{equation}
\widetilde{\delta}_{\nu}(s, \textbf{k}) =  \frac{\widetilde{\delta}_{\nu}(s, k)}{\widetilde{\delta}_{c+b}(s, k)} \widetilde{\delta}_{c+b}(s, \textbf{k}).
\end{equation}

Then the total over-density field is corrected as 

\begin{equation}
\widetilde{\delta}_{t}(s, \textbf{k}) = (1-f_{\nu})\widetilde{\delta}_{c+b}(s, \textbf{k}) +  f_{\nu} \widetilde{\delta}_{\nu}(s, \textbf{k}). 
\end{equation}

Here $f_{\nu} \equiv \bar{\rho}_{\nu} / (\bar{\rho}_{\nu} + \bar{\rho}_{c+b})$ is also carefully treated at each time step, considering that neutrinos are not yet fully non-relativisitc at higher redshift. Eventually the original $\widetilde{\delta}_{c+b}(s, \textbf{k})$ in \texttt{Gadget2} is substituted by this corrected $\widetilde{\delta}_{tot}(s, \textbf{k})$. In this way the gravitational potential and thus the long-range force on CDM simulation particles are corrected with neutrino free-streaming effect.

7. $\widetilde{\delta}_{\nu}(s, k)$, $\widetilde{\delta}_{c+b}(s, k)$ and the scale factor $a_n$ for this time step are saved as the initial conditions for the next PM calculation.

8. Steps 4-7 are repeated every time when calculation of the PM force is called in the \texttt{Gadget2} simulation, until redshift 0 when the simulation ends. At all the designated redshifts for output, $\widetilde{\delta}_{\nu}(s, k)$ are also recorded.

This simulation scheme we adopt is very similar to that of \cite{ab2013}, except for a few minor differences:

1. According to the description in \cite{ab2013}, the authors use the same relative amplitude of over-density for both CDM (and baryons) and neutrinos in the initial condition, while we generate the initial power spectra of CDM and massive neutrinos separately from \texttt{CAMB}, and set their different initial over-density fields accordingly. 

2. In the calculation of linear evolution of neutrino over-density, \cite{ab2013} includes both the 0th order (over-density) and 1st order (bulk velocity) terms in the multipole moment expansion, while we only include the 0th order. However, since the bulk velocity is much less than the thermal velocity, which is the source of the 0th order term, it is safe to neglect it. 

Since the directly measured power spectrum from a snapshot is effectively $P_{c+b}(k)$, the same weighted average Eq.(3.6) is applied again to obtain the total power spectrum. We use our own code for the power spectrum measurement, using the Cloud-in-Cell scheme for the density field distribution. Jing's power-law iterative correction for the window effect is also implemented for the final result \cite{ypjing}.

In \texttt{Gadget2} simulations, the calculation of the PM force is not strictly applied at every time step. In between two PM steps, the time interval is sometimes further divided into shorter steps, during which some of the simulation particles would be drifted according to their velocities. This should not affect the consistency of our correction on the PM force, as long as the corresponding scale factor $a$ is correctly recorded. 

\subsection{Particle-based neutrino simulation}

In our particle-based neutrino simulation, we have $N^3$ CDM particles and another $N^3$ neutrino particles. So the masses and softening lengths for CDM and neutrino particles are scaled according to the direct ratio of $\Omega_{c+b}$ and $\Omega_{\nu}$. The major modification lies in the initial condition generator \texttt{2LPT}, since the dynamics in \texttt{Gadget2} remains the same with an extra kind of particles. Firstly, for the initial configuration, each neutrino particle is displaced from a CDM particle by half the average distance $L/2\sqrt[3]{N}$ in the positive direction of each axis. The spatial perturbations of both kinds of particles are then applied according to their \texttt{CAMB} power spectra respectively. Secondly, for the velocity field, apart from the gravitational flow which is related to the spatial perturbation, the neutrino thermal velocity is included, which actually dominates. The momenta of thermally produced cosmological neutrinos follow the Fermi-Dirac distribution Eq.(2.1), with $\xi=\mu/T = 0$, as the particle-based method cannot be used to simulate neutrinos with finite chemical potential. So the probability of a neutrino particle's momentum to be smaller than $p_n$ is

\begin{equation}
P_n(p'<p_n) = \frac{\int _0^{p_n} f(p')p'^2dp'}{\int _0^{\infty} f(p')p'^2dp'}.
\end{equation}

Then the magnitude of the thermal velocity $|\textbf{v}_{th}|$ is calculated from the momentum $p$, sampled according to (3.7). Since the typical velocity is of order $0.1c$, special relativistic effect is included in evaluating $|\textbf{v}_{th}|$. The thermal velocity needs to be further divided by $\sqrt{a}$ to be consistent with the internal unit of \texttt{Gadget2}. Uniform distribution of spherical angles $\phi$ and $\sin\theta$ is also applied to $\textbf{v}_{th}$, so that there is no preference in directions.

\subsection{Comparison between grid-based and particle-based simulations}

We check the validity of our grid-based simulation code by comparing it with the particle-based simulation. The ratios of the total power spectra of these two neutrino simulations relative to a pure CDM simulation $R(k) \equiv P_{\nu, cdm}(k)/P_{cdm}(k)$ are compared. Since the particle-based simulation has $N^3$ CDM and neutrino particles each, for fairness we also require $2N^3$ CDM particles for the grid-based simulation and the pure CDM simulation, with the same unperturbed initial configuration and mass splitting. We choose $N=128$ through out this work. Please note that in the setup of the $2N^3$ CDM particles, they are under different categories of simulation particles, having different masses according to $\Omega_c / \Omega_\nu$ (same as in the particle-based simulation). In this way we have the same mass and spatial resolution when comparing the results.

We use the same cosmological parameter set in this comparison, where $\Omega_m=\Omega_{\nu} + \Omega_{c+b} =0.3$, $\Omega_b = 0.05$, $\Omega_{\Lambda} = 0.7$, $h = 70$, the spectral index $n_s = 1$, and the amplitude of the primorial power spectrum $A_s = 2.43(*10^{-9})$. This set of cosmological parameters is the same as in \cite{ab2013}, the result of which is also a reference for our comparison. The simulation box size is $200$ $h^{-1}\textrm{Mpc}$, and the starting redshift is $z=49$.

\begin{figure}
\subfigure[Fig. 2a]{
\includegraphics[width=8cm, height=7cm]{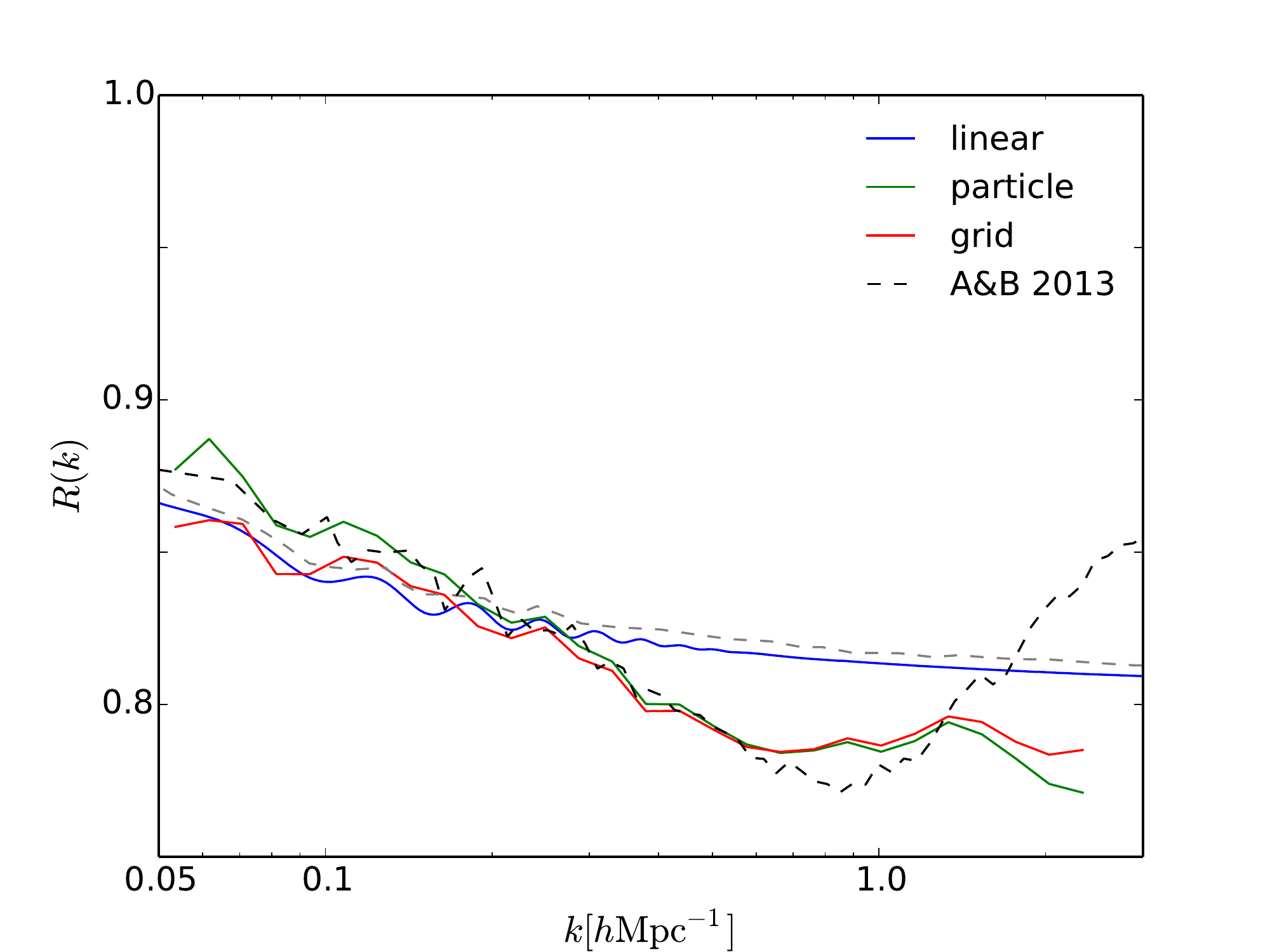}
}
\subfigure[Fig. 2b]{
\includegraphics[width=8cm, height=7cm]{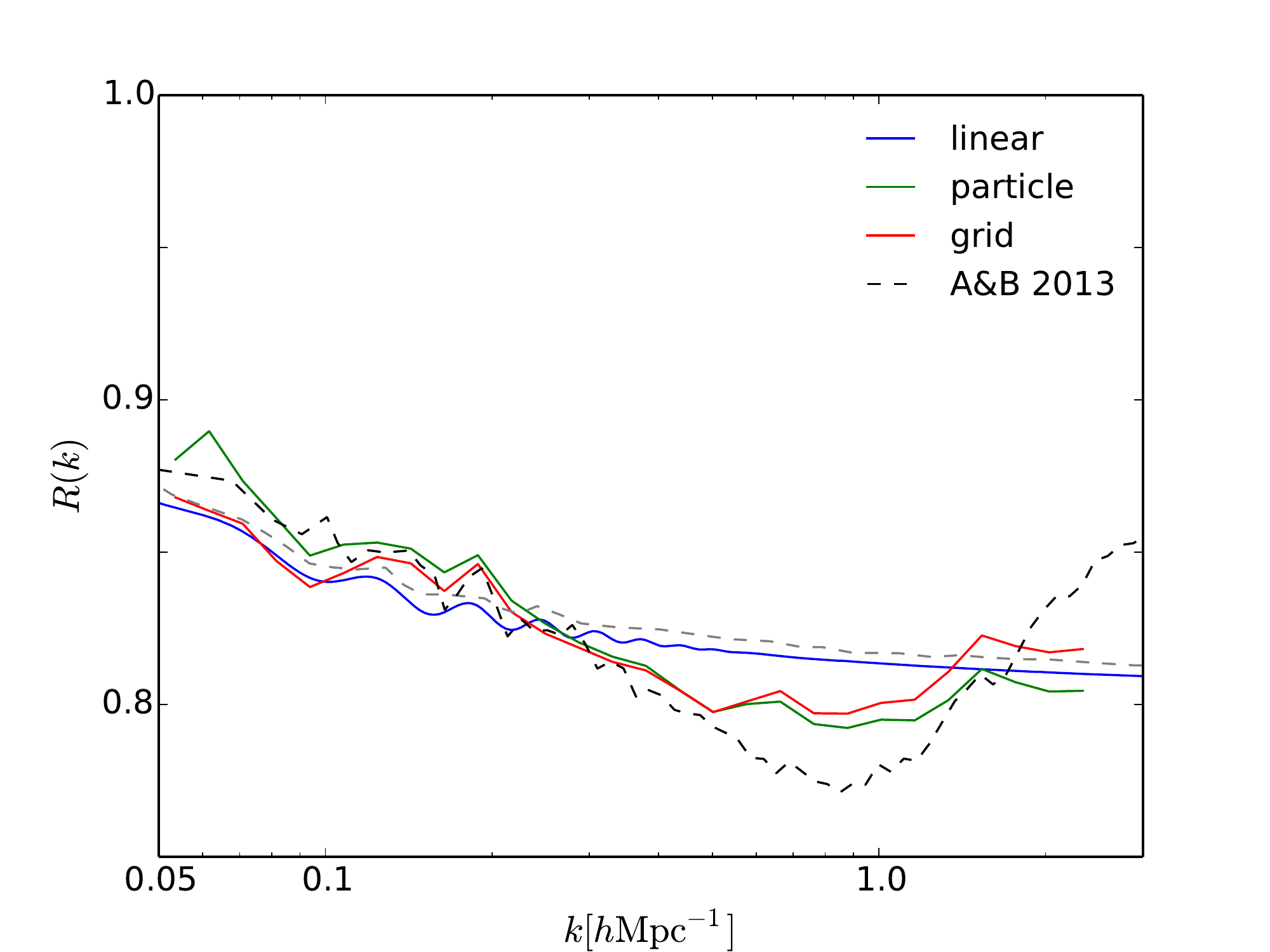}
}
\subfigure[Fig. 2c]{
\includegraphics[width=8cm, height=7cm]{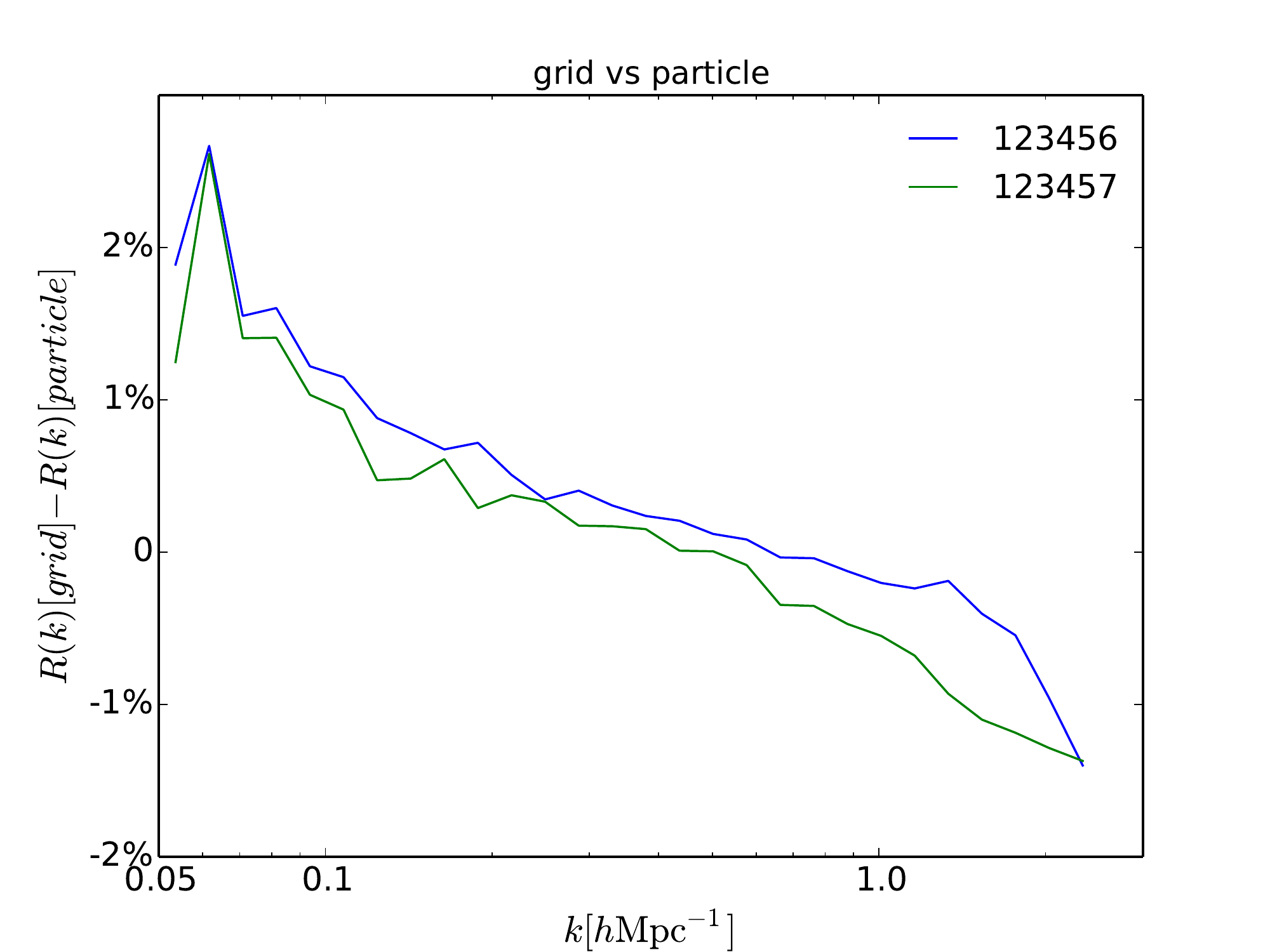}
}

\caption{\label{fig2} a) Relative power spectra $R(k)$ of our grid-based (solid red), particle-based (solid green) simulations and the linear result by \texttt{CAMB} (solid blue). The dashed black and grey curves are reference curves from \cite{ab2013}, for non-linear and linear power spectra respectively. The random seed for the initial condition is $123456$. b) Same as Fig. 2a, but the random seed is $123457$. c) Discrepancy between relative power spectra from grid-based and particle-based simulations. The blue and green curves are for seeds $123456$ and $123457$ respectively. }
\end{figure}

Fig. 2a and Fig. 2b show the relative power spectra of grid-based and particle-based simulations, with different realizations. We can see from Fig. 2a that the results of both the grid-based and particle-based methods are in good agreement with \cite{ab2013}. However, as seen in Fig. 2b, the results for both methods differ from the reference curve of \cite{ab2013} by a small amount. This is due to cosmic variance as we have only changed the random seed for initial conditions of simulations in Fig. 2a and Fig. 2b.  Nevertheless, the grid-based and particle-based methods still produce consistent results. As shown in Fig. 2c, the difference of $R(k)$ between these two simulation methods $R(k)[grid] - R(k)[particle]$ is within $0.5\%$ in the wave-mode range of $0.2 \sim 1.5$ $h\textrm{Mpc} ^{-1}$.

Compared to the linear prediction by \texttt{CAMB}, the relative power spectrum in N-body simulation shows an extra suppression for $k$ from 0.2 to 2  $h\textrm{Mpc} ^{-1}$. References \cite{sb11} and \cite{ab2013} give a qualititive explanation to this behaviour, that the neutrino free-streaming effect firstly delays the the start of the non-linear growth, and secondly pushes the non-linear scale to a larger wave number. This trough in the power spectrum suppression relative to that of CDM is a special signature for N-body simulations including massive neutrinos. Quantitatively the scoop shape can be understood via the halo model. As mentioned in \cite{scoop_halo_model}, the strongest suppression on the matter power spectrum is at $k\sim 1 h \textrm{Mpc}^{-1}$, which is the typical transition scale of the 1-halo and 2-halo terms in the halo model. The $P_k$ in this scale is dominated by the 1-halo term of the most massive haloes. \cite{halo_mass_func, brandbyge_halo, zzc_thesis} show that in the cosmology with massive neutrinos, the abundance of massive haloes are more affected than small haloes, which leads to the signature scale of the strongest suppression. This can be explained via the reduced variance in the Press-Schechter halo mass function, which is for the spehrical collapse model of isolated haloes, and via the hierachical formation of large haloes \cite{zzc_thesis}.

\section{Results}

\subsection{Neutrino effects on the power spectrum}

In this paper all three mechanisms for the cosmic neutrinos to affect the large-scale structure are consistently considered: the free-streaming effect, the modified expansion history and the refitting of the cosmological parameters with $m_{\nu}$ and $\eta$. The effect of the modified background expansion is only about $1\%$, as shown in Section [2.4]. So we do not discuss it separately, but implement it in the grid-based free-streaming code. Then we analyze the effects of $m_\nu$ and $\eta$ separately, by controlling input variables. Six sets of simulations are generated, the parameters of which are listed in Table 2. Our grid-based simulations have $128^3$ CDM particles, box size of $200$ $h^{-1}\textrm{Mpc}$ on each side and starting redshift of 49.

\begin{table}[tbp]
\centering
\begin{tabular}{ |c  c | c | c |c |c  c  c c | c  c |}
\hline
no. & model & $m_{\nu}$(eV) & $\eta$ & $H_0$ & $\Omega_{b}$ &$\Omega_{cdm}$ & $\Omega_{\nu}$ & $\Omega_{\Lambda}$ & $A_s (10^{-9})$ & $n_s$\\
\hline
B1 &  fiducial & 0 & 0 & 67.74 & 0.0484 & 0.2613 & $\sim 10^{-5}$ & 0.6903 & 2.204 & 0.9641\\
B2 & fs &0.048 & 0.0 & 67.74 & 0.0484 & 0.2580  & 0.0033 & 0.6903 & 2.204 & 0.9641\\
B3 & fs &0.048 & 0.359 & 67.74 & 0.0484 & 0.2579  & 0.0034 & 0.6903 & 2.204 & 0.9641\\
\hline
B4 & fs, [all] &0.048 & 0.359 & 66.95 & 0.0496 & 0.2707 & 0.0035 & 0.6762 & 2.240 & 0.9660\\
B5 & fs, $[\Omega_b, \Omega_c, H_0]$& 0.048 & 0.359 & 66.95 & 0.0496 & 0.2707 & 0.0035 & 0.6762 & 2.204 & 0.9641\\
B6 & [all] & 0.0 & 0.0 & 66.95 & 0.0496 & 0.2707 & $\sim 10^{-5}$ & 0.6797 & 2.240 & 0.9660\\
\hline
\end{tabular}
\caption{\label{table 2} Parameters of N-body simulations. The data set B1 is the pure CDM simulation. In data sets B2 and B3 only the effect of neutrino free-streaming is considered, with and without the neutrino degeneracy parameter $\eta$ respectively. For data set B4, we include the refitting of all cosmological parameters, while for B5, only $\Omega_i$ and $H_0$ are refitted, but not $A_s$ or $n_s$. In the data set B6, all cosmological parameters are chosen to be the same as those in B4, but the neutrino free-streaming effect is not included. }
\end{table}

In Table 2, the data set B1 is a pure CDM simulation ($m_\nu = 0$, $\eta = 0$), while for data sets B2 and B3 the neutrino free-streaming effect is included. In B2 we only have massive neutrinos of $m_{\nu} = 0.048$ eV ($\eta = 0$), while in B3 we add $\eta = 0.359$ as well. These are the mean values from our  \texttt{CosmoMC} fitting of the Planck data. We follow the conventional treatment to keep $\Omega_m = \Omega_c + \Omega_b + \Omega_{\nu}$ a fixed value in data sets B2 and B3, and so the cosmological parameters used in B2 and B3 are the same as those in B1, except for a small redistribution of density from $\Omega_{cdm}$ to $\Omega_\nu$.  In data sets B4 and B5, the refitted cosmological parameters are used together with the neutrino free-streaming. In the data set B4, all the affected cosmological parameters are modified accordingly, including $\{\Omega_i\}$, $H_0$, the scalar amplitude $A_s$ for primordial perturbation and the spectral index $n_s$. In the data set B5 we only include the refitting of $H_0$ and $\{\Omega_i\}$. The data set B6 is another CDM simulation where only the refitted cosmological parameters are included, but not the neutrino free-streaming. The comparison of these cases  are shown in Fig. 3a. We also show the corresponding linear power spectra given by \texttt{CAMB}, in dashed curves.

\begin{figure}
\subfigure[Fig. 3a]{
\includegraphics[width=8cm, height=7cm]{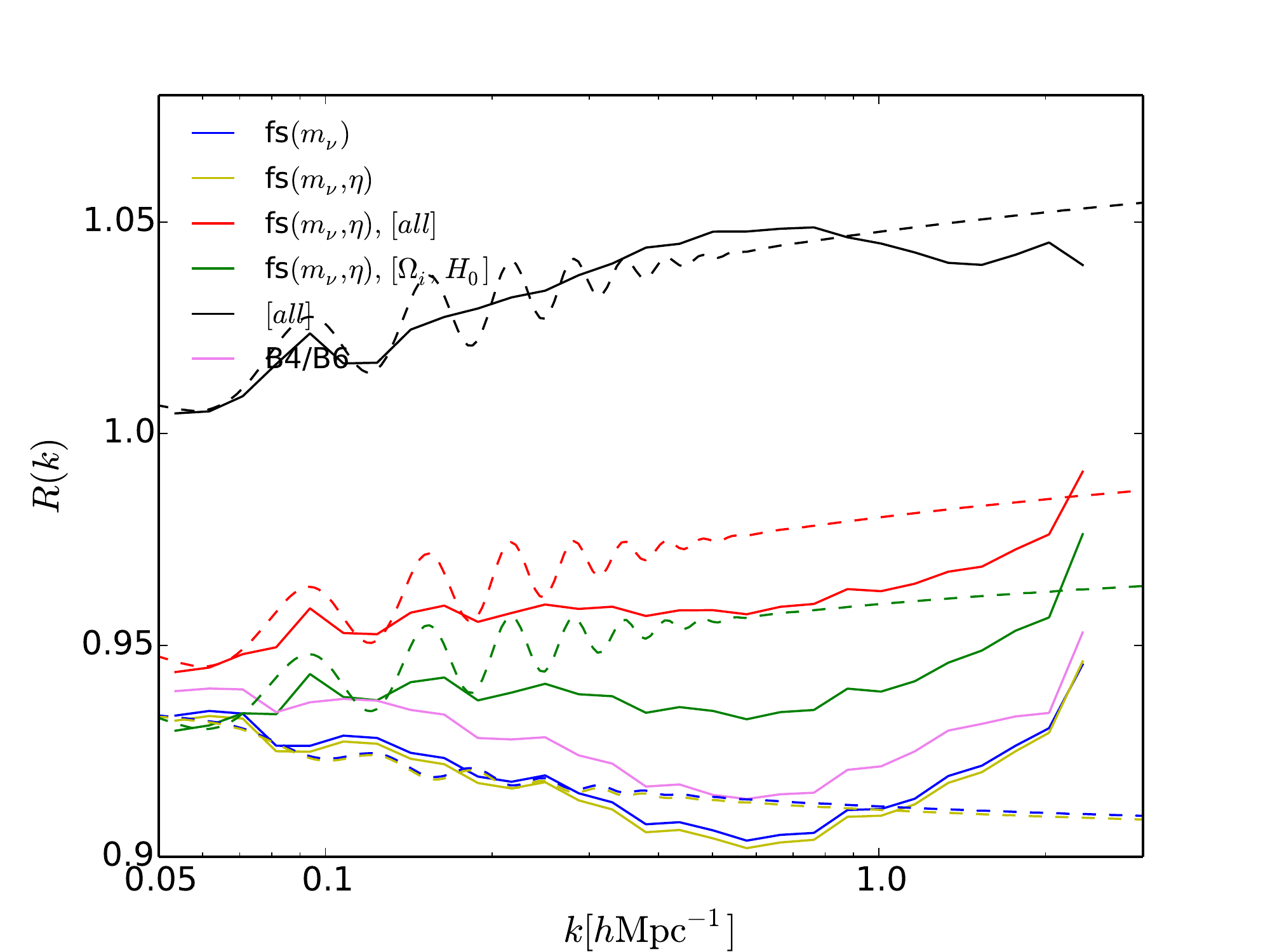}
}
\subfigure[Fig. 3b]{
\includegraphics[width=8cm, height=7cm]{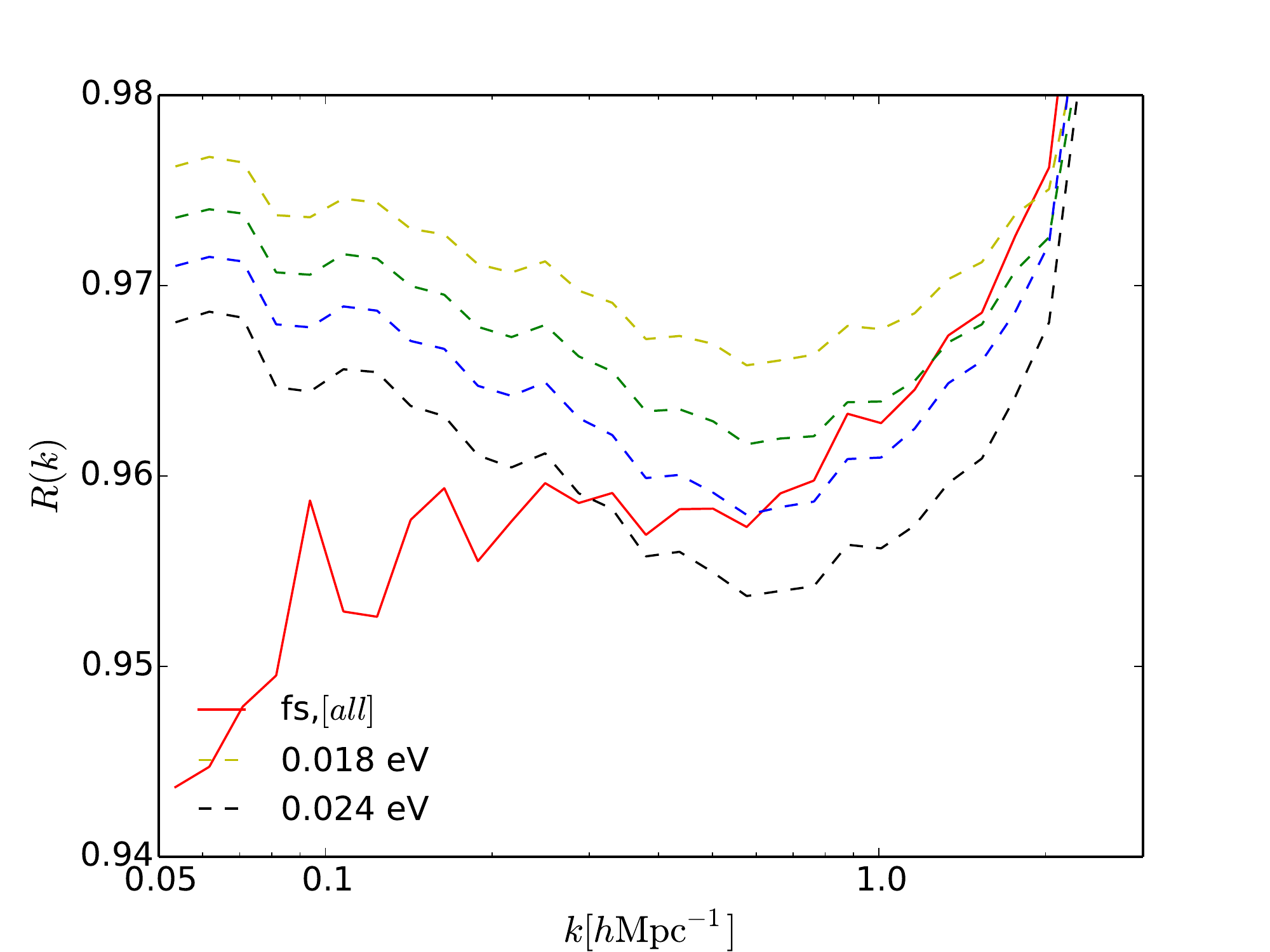}
}

\caption{\label{fig3} a) Relative power spectra $R(k) \equiv P(m, \xi, k) / P(0,0, k)$ for data sets in Table 2. The solid blue, yellow, green, red and black curves show N-body simulation results of data sets B2, B3, B4, B5, B6 respectively, relative to B1, while the dashed ones are corresponding ratios of linear power spectra given by \texttt{CAMB}. The solid violet curve shows the ratio of B4/B6, which reflects the pure free-streaming effect in the new cosmology. b) Comparison between the relative power spectrum using the mean values of  \texttt{CosmoMC} fit, $m_{\nu} = 0.048$ eV and $\eta = 0.359$ in B4, and relative power spectra of free-streaming-only simulations. The solid red curve shows the result of B4 relative to B1, while the dashed yellow, green, blue and black curves show the free-streaming-only power spectra with $m_{\nu}$ sampled to be $0.018, 0.020, 0.022, 0.024$ eV respectively.}
\end{figure}

Comparing data sets B2 and B3 (blue and yellow) we see that the impact of ${\eta}$ is small, if only the neutrino free-streaming effect is considered. This is because $\Omega_{\nu}$ is only shifted slightly by $\eta = 0.357$ compared to $\eta = 0$. The magnitudes of suppression of $R(k)$ in data sets B2 and B3 are about $10\%$ at most, with the same trough-like deviation from the linear prediction as we have reported in Fig. 2a. This is also consistent with the empirical result from \cite{brandbyge08} that $\Delta P(k)/P(k) \approx -9.8 \Omega_{\nu}/\Omega_m$. But for the data set B5 (green), where we include the refitted $H_0$ and $\{\Omega_i\}$, the suppression is decreased. When we further include the refitted $A_s$ and $n_s$ in the data set B4 (red), the suppression is further decreased to about $4\%$ only. In the data set B6 where we only consider the refitted cosmological parameters, the total power spectrum is even increased by about $5\%$ at most. The difference between data sets $B4$ and $B6$ (black and red) is still about $9\% -10\%$, consistent with previous results. We have also plotted the ratio of B4/B6 as the violet curve, which is the pure free-streaming effect with the newly fitted cosmological parameters. As expected, it is pretty close to the blue curve (B3/B1), showing the robustness of free-streaming effect in different cosmologies. Therefore, we can conclude from Fig. 3a that the refitted cosmological parameters increase the total power spectrum, while the free-streaming effect and the modified expansion rate together result in a suppression. A completely consistent treatment of cosmological neutrinos' impact on the matter power spectrum must include all these mechanisms. Similar results are also observed in a recent paper \cite{jliu}, where the cosmological parameters are allowed to vary, apart from the neutrino free-streaming effect.

We further conduct a degeneracy test to see what value of $m_{\nu}$  would a 4\% suppression in the matter power spectrum (data set B4, red curve in Fig. 3a) imply, if only the free-streaming of massive neutrinos is considered ($\eta = 0$, no refitting of cosmological parameters). From \cite{brandbyge08}, $\Delta P(k) / P(k) \approx -9.8 \Omega_{\nu} / \Omega_m$, and so we estimate the neutrino mass to be $m_{\nu} \approx 0.022$ eV. Then we sample four sets of $m_{\nu} = [0.018, 0.020, 0.022, 0.024]$ eV and ${\eta} = 0$, and generate simulations accordingly. In Fig. 3b we can see that the closest estimate is $m_{\nu} = 0.02$\ eV, while the shape of the suppression curve is slightly different. This is to say, the suppression of the matter power spectrum considering only the free-streaming effect of neutrinos with $m_{\nu} = 0.022$ eV is roughly the same as that with $m_{\nu} = 0.048$ eV and $\eta = 0.359$, with refitting of cosmological parameters consistently included. Thus the conventional method to constrain the neutrino mass solely from the free-streaming effect can have an error as large as $100\%$, if the finite chemical potential and refitting of cosmological parameters are ignored.

It is clear that the free-streaming effect mainly depends on the neutrino mass, while the effects of $m_{\nu}$ and $\eta$ on the refitted cosmological parameters are still to be unveiled. We will try to break this degeneracy in the following section.

\subsection{Results for varying $m_{\nu}$ and $\eta$}

Following the spirit of Section 4.1, we sample a series of fixed parameters $\{m_{\nu}, \eta\}$, put each of them into our MCMC refitting to obtain the corresponding cosmological parameters, and generate grid-based neutrino simulations with our modified \texttt{Gadget2} code accordingly. The fitting results of cosmological parameters are listed in Table 3. For the convenience of future use, we numerically fit the changes of corresponding cosmological parameters (including their uncertainties from the \texttt{CosmoMC} refitting) as functions of $m_{\nu}$ and $\eta$ as in Eqs.(4.1). Note that here the cosmological parameters are normalized by their values in the $\Lambda$CDM cosmology and $m_{\nu}$ is normalized by $0.1$ eV.

 \begin{equation}
\begin{split}
 \frac{\Delta H_0}{H_{0[\Lambda \mathrm{cdm}]}} [\%] = -4.40 \frac{m_{\nu}}{0.1\ \textrm{eV}} + 4.50 \eta ^2 [\%]					\\
\frac{\Delta \Omega_b}{\Omega_{b[\Lambda \mathrm{cdm}]}}  [\%] = 8.62 \frac{m_{\nu}}{0.1\ \textrm{eV}}  -7.42 \eta ^2 [\%]		\\
\frac{\Delta \Omega_c}{\Omega_{c[\Lambda \mathrm{cdm}]}}  [\%] = 9.67  \frac{m_{\nu}}{0.1\ \textrm{eV}}  - 4.07 \eta ^2  [\%]			\\
 \frac{\Delta A_s}{A_{s[\Lambda  \mathrm{cdm}]}}  [\%]= 1.98  \frac{m_{\nu}}{0.1\ \textrm{eV}} + 3.67 \eta ^2  [\%]						\\
 \frac{\Delta n_s}{n_{s[\Lambda \mathrm{cdm}]}}  [\%]= -0.21  \frac{m_{\nu}}{0.1\ \textrm{eV}} + 1.62 \eta ^2  [\%]
 \end{split}
 \end{equation} 

In our least squares regression to obtain these numerical formulae, we keep $m_{\nu}$ in first order, while $\eta$ in second order. This is because the regression result shows that the coefficient of ${\eta^1}$ is 0 within uncertainty. So we simply omit the first order term. This is also physically reasonable, as in the calculation of the neutrino energy density in Eq. (2.9), the contributions of neutrinos and anti-neutrinos cancel at the first order in $\eta$.

\begin{table}[tbp]
\centering
\begin{tabular}{ c | c | c |c |c  c c  c| c  c | c }
\hline
no. & $m_{\nu}$(eV) & $\eta$ & $H_0$ & $\Omega_{b}$ &$\Omega_{cdm}$  & $\Omega_{\nu}$ & $\Omega_{\Lambda}$ & $A_s (10^{-9})$ & $n_s$ & $\Delta R(\%)$\\
\hline
C0 & 0.0 & 0.0 & 67.74 & 0.0484 & 0.2613  & $\sim 10^{-5}$ & 0.6903 & 2.204 & 0.9641 & 0.0\\
\hline
C1 & 0.02 & 0.0 & 67.27 & 0.0491 & 0.2651  & 0.0014 & 0.6844 & 2.209 & 0.9638 & -2.95 $\pm$ 5.41 \\
C2 & 0.02 & 0.251 & 67.46 & 0.0489 & 0.2644  & 0.0014 & 0.6853 & 2.218 & 0.9647 & -1.76 $\pm$ 5.41\\
C3 & 0.02 & 0.502 & 68.00 & 0.0482 & 0.2628  & 0.0014 & 0.6876 & 2.230 & 0.9675 & 1.44 $\pm$ 5.58 \\
C4 & 0.02 & 0.754 & 68.97 & 0.0471 & 0.2596  & 0.0014 & 0.6919 & 2.255 & 0.9725 & 7.51 $\pm$ 6.01 \\
C5 & 0.02 & 1.006 & 70.39 & 0.0456 & 0.2548  & 0.0015 & 0.6981 & 2.292 & 0.9797 & 16.63 $\pm$ 6.68 \\
\hline
C6 & 0.05 & 0.0 & 66.32 & 0.0504 & 0.2733  & 0.0036 & 0.6727 & 2.223 & 0.9633 & -7.55 $\pm$ 5.05\\
C7 & 0.05 & 0.251 & 66.48 & 0.0502 & 0.2729  & 0.0036 & 0.6733 & 2.229 & 0.9639 & -6.50 $\pm$ 5.12 \\
C8 & 0.05 & 0.502 & 67.04 & 0.0496 & 0.2708  & 0.0037 & 0.6759 & 2.246 & 0.9670 & -3.50 $\pm$ 5.33\\
C9 & 0.05 & 0.754 & 67.96 & 0.0485 & 0.2679  & 0.0037 & 0.6799 & 2.272 & 0.9719 & 2.12 $\pm$ 5.68\\
C10 & 0.05 & 1.006 & 69.35 & 0.0468 & 0.2630  & 0.0038 & 0.6864 & 2.308 & 0.9790 & 10.52 $\pm$ 6.19 \\
\hline
C11 & 0.08 & 0.0 & 65.39 & 0.0518 & 0.2816  & 0.0060 & 0.6606 & 2.237 & 0.9625 & -12.68 $\pm$ 4.81\\
C12 & 0.08 & 0.251 & 65.56 & 0.0516 & 0.2811  & 0.0060 & 0.6613 & 2.244 & 0.9634 & -12.55 $\pm$ 4.79\\
C13 & 0.08 & 0.502 & 66.11 & 0.0509 & 0.2789  & 0.0060 & 0.6642 & 2.260& 0.9663 & -8.80 $\pm$ 4.98\\
C14 & 0.08 & 0.754 & 66.97 & 0.0498 & 0.2764  & 0.0061 & 0.6677 & 2.287 & 0.9713 & -3.56 $\pm$ 5.30\\
C15 & 0.08 & 1.006 & 68.25 & 0.0482 & 0.2723  & 0.0062 & 0.6733 & 2.322 & 0.9781 & 3.76 $\pm$ 5.87 \\
\hline
C16 & 0.048 & 0.359 & 66.95 & 0.0496 & 0.2707 & 0.0035 & 0.6762 & 2.240 & 0.9660 & -4.06 $\pm$ 7.08\\
\hline
\end{tabular}
\caption{\label{table 3} Sampled values of $m_{\nu}$ and $\eta$, and results of the refitted cosmological parameters. The data set C0 is the pure $\Lambda$CDM case. In groups C1-C5, C6-C10 and C11-C15, $m_{\nu}$ are $0.02, 0.05$ and $0.08$ eV respectively, while within each group, $\eta$ is sampled to be $[0.0, 0.251, 0.502, 0.754, 1.006]$ in sequence. The data set C16 is the mean values of our  \texttt{CosmoMC} fit, with both $m_{\nu}$ and $\eta$ freely varying. Other groups of sampled values are selected to be around this mean-value set. The last column is the measurement of the percentage difference $\Delta R$ of the averaged power spectra at $z=0$ in the $k$-range $[0.3, 1.0] h \mathrm{Mpc}^{-1}$, which we select to be a convenient quantity for comparison of different simulations.}
\end{table}

From Eq.(4.1), we can see that for the fractional changes in $H_0$, $\Omega_b$ and $\Omega_c$, the coefficients of $m_{\nu}$ and $\eta^2$ have opposite signs, and are comparable in magnitude. For the scalar amplitude, both $m_{\nu}$ and $\eta^2$ lead to an increased $A_s$. As for the power spectral index $n_s$, the coefficient of $\eta^2$ is almost one order larger than that of $m_{\nu}$. All these cosmological parameters are shifted by  $1\%-10\%$, which are significant. Thus noticeable effects of $m_{\nu}$ and $\eta^2$ on the matter power spectrum can also be expected.

In Fig. 4a we select four cases with $m_{\nu} =[0.02, 0.08]$ eV and $\eta = [0.0, 1.006]$, together with their mean values $m_{\nu} = 0.048$ eV, $\eta = 0.359$ in our refitting of the Planck data (C1, C5, C11, C15 and C16 in Table 3) to plot the ratios of power spectra to that of $\Lambda$CDM cosmology (C0 in Table 3). The cases with $\eta =0$ restore the typical suppression on the power spectrum due to the free-streaming effect, while a non-zero value of $\eta$ leads to an enhancement of the power spectrum. Thus the final power spectrum is the result of these two competing factors. We can also see from this figure that the effects of $m_{\nu}$ and $\eta^2$ are comparable in magnitude for the range of parameter values we considered. 

\begin{figure}
\subfigure[Fig. 4a]{
\includegraphics[width=8cm, height=7cm]{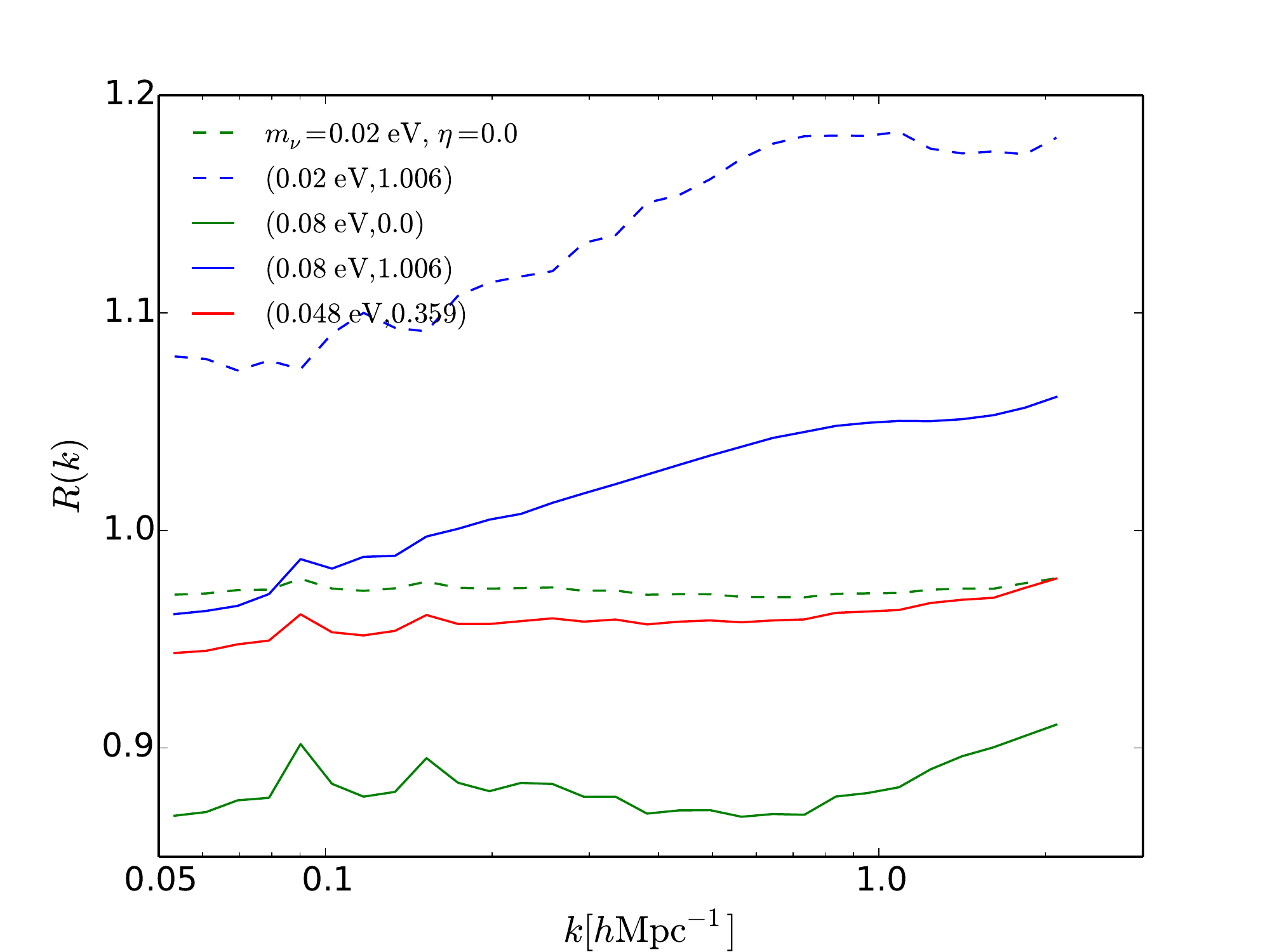}
}
\subfigure[Fig. 4b]{
\includegraphics[width=8cm, height=7cm]{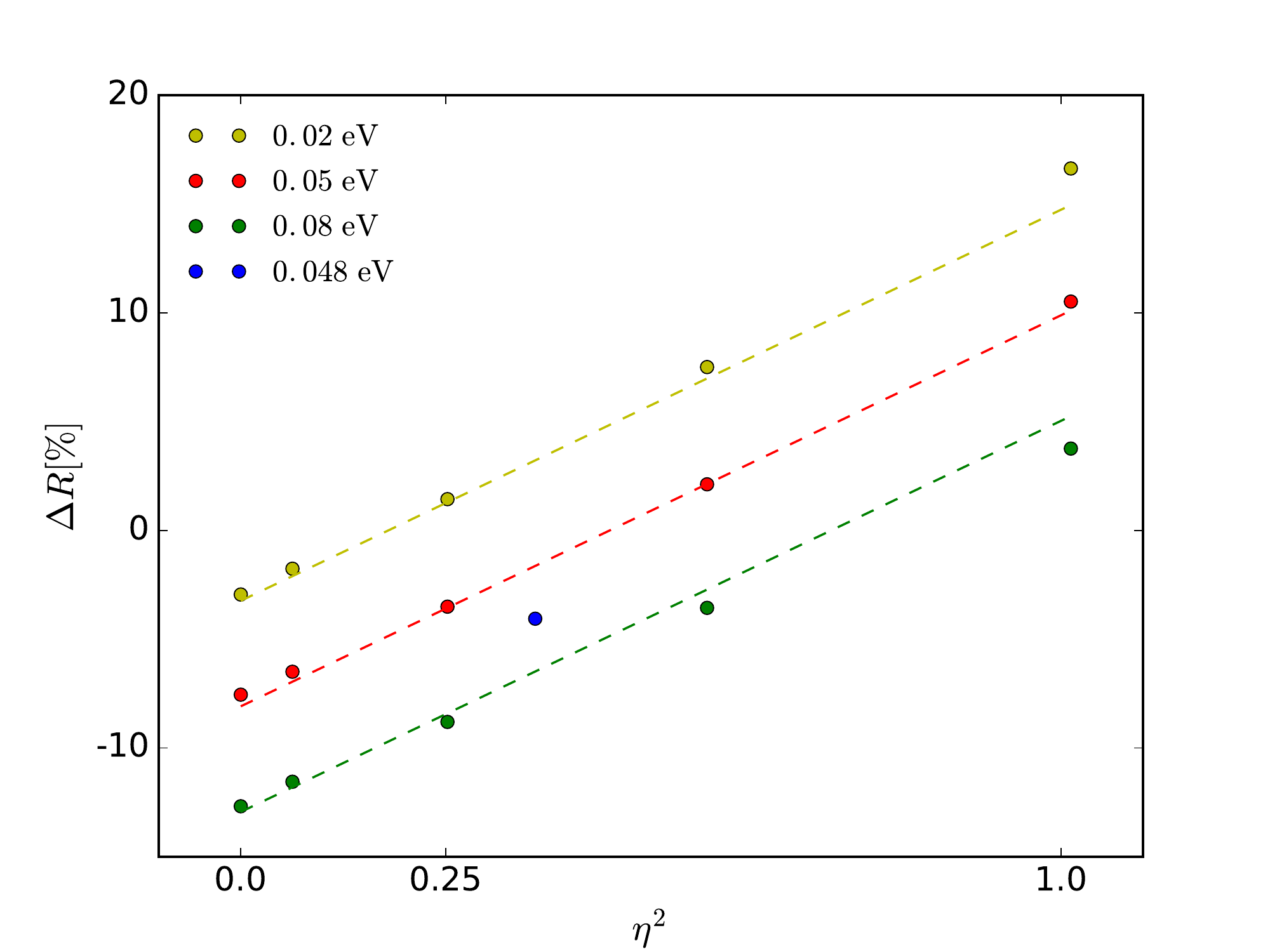}
}
\subfigure[Fig. 4c]{
\includegraphics[width=8cm, height=7cm]{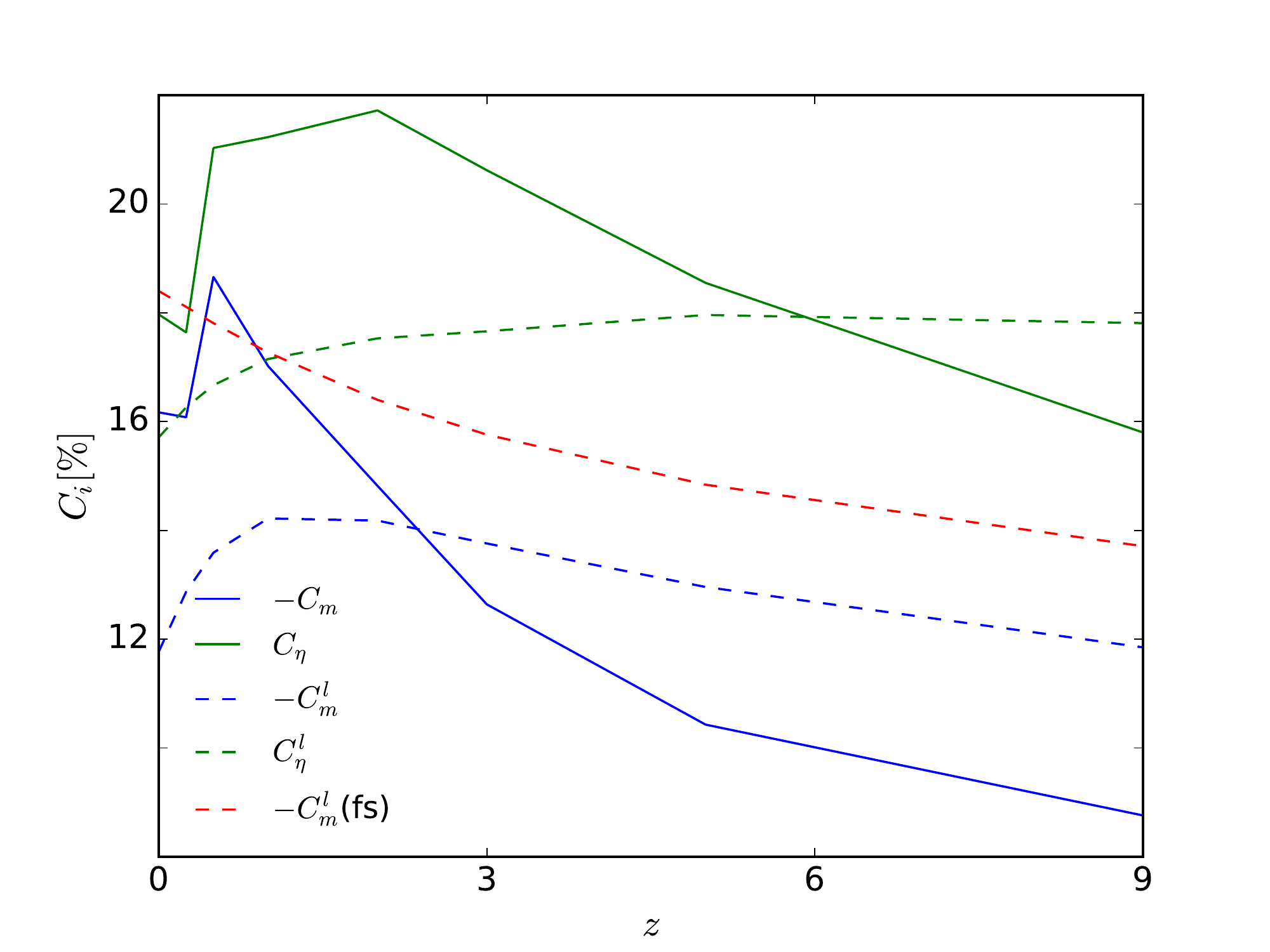}
}
\subfigure[Fig. 4d]{
\includegraphics[width=8cm, height=7cm]{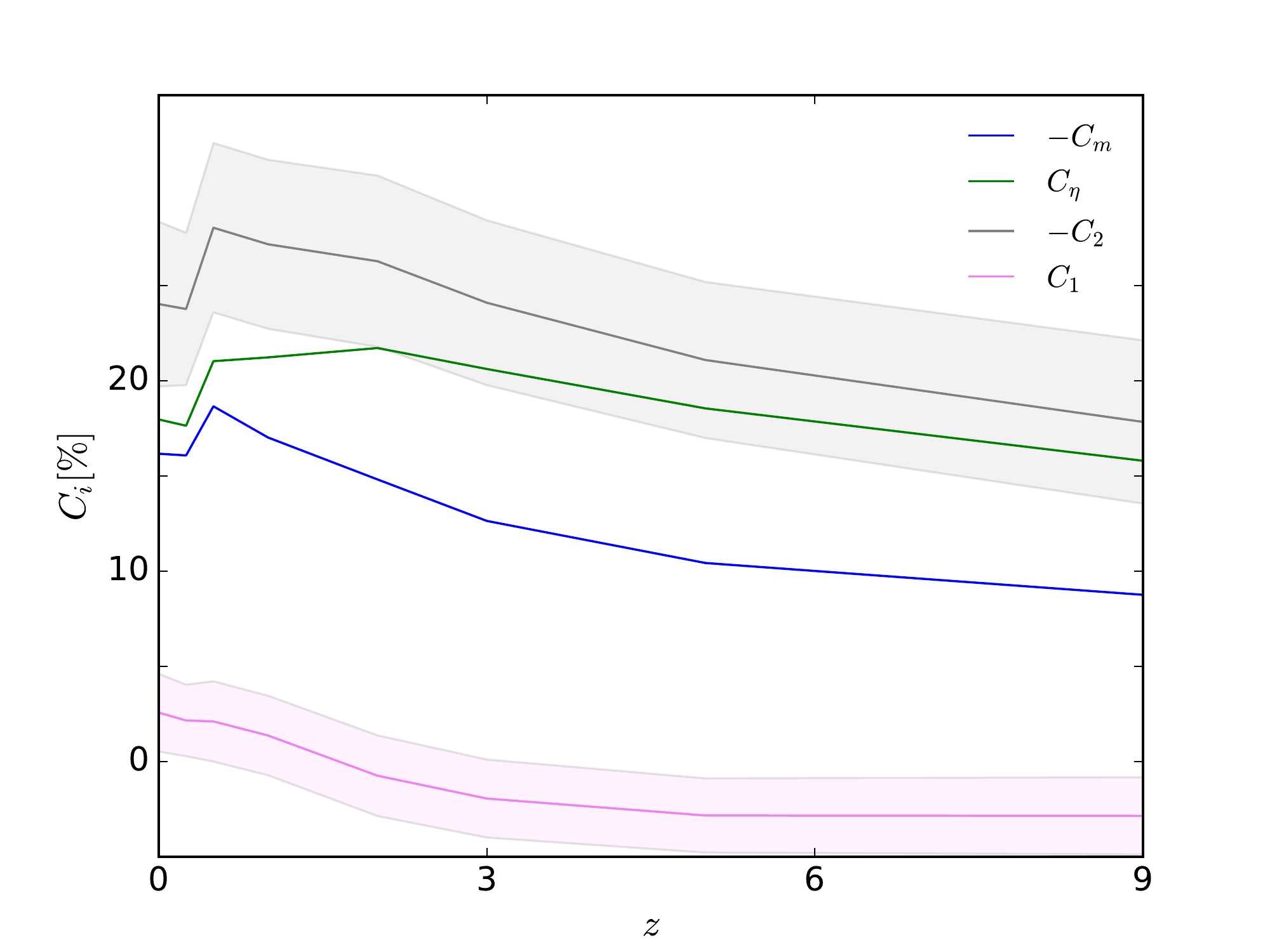}
}
\caption{\label{fig4} a) Relative power spectra $R(k)$ for data sets C1 (dashed green) and C5 (dashed blue), where $m_{\nu} = 0.02$ eV and $\eta = 0.0, 1.006$ respectively, data C11 (solid red) and C15 (solid blue), where $m_{\nu} = 0.08$ eV and $\eta = 0.0, 1.006$ respectively, and data C16 (solid red), where the mean values of refitted cosmological parameters are used. b) Percentage difference $\Delta R$ of the averaged power spectra in the $k$-range $[0.3, 1.0]\ h\textrm{Mpc}^{-1}$, for the 16 data sets in Table 2 from that of the $\Lambda$CDM (C0). Yellow, red and green dots show groups C1-C5,  C6-C10 and C11-C15, where $m_{\nu} = 0.02, 0.05$ and $0.08$ eV respectively. The blue dot is for the mean-value data set C16. The dashed curves are reference fittings from Eq.(4.3). c) Redshift evolution of coefficients $-C_m$ and $C_\eta$ from the fitted formula Eq.(4.3). The solid green and blue curves show $-C_m$ and $C_\eta$ from grid-based neutrino simulations, while the dashed green and blue curves show $-C_m$ and $C_\eta$ of the linear power spectra given by \texttt{CAMB}. The dashed red curve shows the $-C_m$ of the linear power spectrum when only neutrino free-streaming is considered. d) Redshift evolution of coefficients $-C_m$ (blue), $C_\eta$ (green), $C_1$ (violet) and $-C_2$ (grey) from Table 4. The shaded regions show the uncertainties of $C_1$ and $-C_2$ respectively.}
\end{figure} 

From Fig. 4a, these relative power spectra are of different shapes and do not have a universal signature. For convenience, we define a physical quantity to characterize the effects of $m_{\nu}$ and $\eta^2$. We measure the averaged value of ratios of power spectra $R(k) \equiv P(m_{\nu}, \eta, k) / P(0, 0, k)$ in the $k$ range of $0.3 \sim 1.0\ h\textrm{Mpc}^{-1}$, $\bar{R}_{[0.3, 1.0]}$, where the non-linear power spectrum deviates significantly from the linear one the most, as shown in Fig. 3a. This $k$-range can be customized for future comparison with different observations. Then we define 
\begin{equation}
\Delta R = \bar{R}_{[0.3, 1.0]}  - 1
\end{equation}
to be the fractional difference between these power spectra and the pure CDM case. The values of $\Delta R$ at redshift 0 are listed in Table 3, and plotted as discrete points in Fig. 4b. The uncertainty of $\Delta R$ is also included, the details of which are discussed in Appendix B. We can see clearly that in agreement with the previous argument for regression, $\Delta R$ is linear in $m_{\nu}$ and $\eta^2$. Then we use the same regression strategy to fit for its dependence on $m_{\nu}$ and $\eta^2$ with the 17 datasets in Table 3. The two-variable linear regression of $\Delta R$ against $m_{\nu}$ and $\eta^2$ gives 

\begin{equation}
\Delta R (z) [\%] = C_m\frac{m_{\nu}}{0.1\ \textrm{eV}} + C_{\eta} \eta^2 [\%], 
\end{equation}
where we define $ C_m$ and $C_\eta$ to be the two coefficients, with $C_m(z = 0) = -16.7$ and $C_\eta (z = 0) = 17.97$. We can see that $m_{\nu} \sim \mathcal{O} (0.1$ eV) and $\eta^2 \sim \mathcal{O}(1)$ would have comparable effects on the total matter power spectrum at $z=0$, but in opposite directions.

\begin{table}[tbp]
\centering
\begin{tabular}{ |c | c c| c c c| c|c  c| }
\hline
$z$ & $C_m(\%)$ & $C_\eta(\%)$ & $\sigma _{mm}$& $\sigma_{\eta\eta}$ & $\sigma_{m\eta}$ &$\theta$ ($\pi$) &$C_1(\%)$ & $C_2 (\%)$ \\
\hline
0 & -16.17 & 17.97 & 9.18& 13.75 & -6.93 & 1.80 & 2.57$\pm$2.04 & -24.03 $\pm$ 4.33\\
0.25 & -16.08 & 17.64 & 8.10 & 11.46 & -6.00 &1.80 &2.16$\pm$1.88 & -23.77$\pm$4.00 \\
0.5 & -18.66 & 21.03 & 10.09 & 14.17  & -7.41 & 1.79 &2.11$\pm$2.11 & -28.04$\pm$4.45\\
1 & -17.02 & 21.23 & 9.63 & 14.44 &  -7.26 & 1.80 &1.37$\pm$2.09 & -27.17$\pm$4.44\\
2 & -14.82 & 21.72 & 9.93 & 14.81 & -7.46 & 1.80 &-0.74$\pm$2.12 & -26.28$\pm$4.50\\
3 & -12.64 & 20.62 & 9.26 &  13.70 & -6.92 & 1.80 &-1.94$\pm$2.05 & -24.10$\pm$4.33\\
5 & -10.43 & 18.55 & 8.51 & 12.12 & -6.27  & 1.80 &-2.83$\pm$1.95 & -21.09$\pm$4.10\\
9 & -8.76 & 15.80 & 9.57 & 12.99 & -6.93 &1.79 &-2.85$\pm$2.03 & -17.84$\pm$4.29\\
\hline
\end{tabular}
\caption{\label{table 4} Results of the two-variable regression of $\Delta R$ at different redshifts as a linear function of $m_{\nu}$ and $\eta^2$, the elements of the covariance matrices of these fittings $\sigma _{mm}$, $\sigma_{\eta\eta}$ and $\sigma_{m\eta}$, the rotation angles $\theta$ to a new basis in which the covariance matrices are diagonal, and the corresponding coefficients in the new basis.}
\end{table}

\begin{table}[tbp]
\centering
\begin{tabular}{ |c | c c| c c c| c|c  c| }
\hline
 & $C_m(\%)$ & $C_\eta(\%)$ & $\sigma _{mm}$& $\sigma_{\eta\eta}$ & $\sigma_{m\eta}$ &$\theta$ ($\pi$) &$C_1(\%)$ & $C_2 (\%)$ \\
\hline
$\Omega_b$ & 8.62 & -7.42 &0.49 & 0.52 & -0.32 & 1.76 &-1.13$\pm$0.43 & 11.32 $\pm$ 0.91\\
$\Omega_c$ & 9.67 & -4.07 & 1.23 & 1.30 & -0.81 & 1.76 & -4.18$\pm$0.67 & 9.62$\pm$1.44 \\
$H_0$ & -4.40 & 4.50 & 0.69 & 0.81 & -0.48 & 1.77 &0.35$\pm$0.52 & -6.28$\pm$1.11\\
$A_s$ & 1.98  & 3.67 & 7.71 & 9.17 & -5.33 & 1.77 &-3.89$\pm$1.75 & -1.44$\pm$3.72\\
$n_s$ & -0.21 & 1.62 & 0.18 & 0.21 &  -0.12 & 1.77 &-0.92$\pm$0.26 & -1.35$\pm$0.56\\

\hline
\end{tabular}
\caption{\label{table 4} Same analysis results as Table 4, but for standard cosmological parameters. }
\end{table}

The evolution of $C_m$ and $C_\eta$ in cosmic time is also studied. We repeat the fitting of Eq.(4.3) for a series of redshifts $[0, 0.25, 0.5, 1, 2, 3, 5, 9]$. The results of this two-variable linear regression are listed in Table 4 and plotted in Fig. 4c. The corresponding $C_m^l$ and $C_\eta^l$ for the linear power spectra calculated by \texttt{CAMB} are also shown as dashed lines in Fig. 4c. As a comparison, the $C_m^l$ for the linear power spectrum with only neutrino free-streaming effect considered is plotted as the red dashed line, which monotonically increases with time. This is well-expected, because in the same cosmology, the suppression of structure due to neutrino free-streaming is accumulative, and increases as structure grows. However, the $C_m^l$ of the linear power spectrum (blue dashed curve) deviates from the red dashed curve and even shows a drop at low redshift. This is because the $P(m_{\nu}, \eta, k)$ and $P(0, 0, k)$ are from different cosmologies, and so both the initial power spectra and their growths behave differently. We can also see that the $C_m$ and $C_\eta$ fitted from the non-linear simulations deviate significantly from those for linear power spectra, although both have a similar drop in $z$. Thus it is important to insist on using N-body simulation to study the neutrino-included power spectrum. The green and blue solid curves, for the non-linear $C_m$ and $C_\eta$, overlap at low redshift, but deviate from each other at earlier time. Therefore, although the affected power spectrum is an integrated result of the effects of $m_{\nu}$ and $\eta^2$, it may be possible to break this parameter degeneracy by studying the redshift dependence of $C_m$ and $C_\eta$ in future observations.

There may be correlation between $m_\nu$ and $\eta^2$ in the fitting of the changes in cosmological parameters and relative power spectra into Eqs.(4.2) and (4.3). Thus we calculate the covariance matrices of these fittings and list the elements in Table 4 and Table 5, from which we can see that the off-diagonal terms $\sigma_{m \eta}$ are indeed comparable to the diagonal terms $\sigma_{mm}$ and $\sigma_{\eta \eta}$. Therefore, anti-clockwise rotations with angles $\theta$ are applied to the original basis $(m_\nu, \eta^2)$, so that the new covariance matrices are diagonal in the new basis $(v_1, v_2)$. The rotation angles $\theta$ and coefficients in the new basis $C_1$ and $C_2$ are also listed in Table 4 and Table 5, for the fittings of $\bar{R}_{[0.3, 1.0]}$ and cosmological parameters respectively. It can be seen that $\theta$ for the fitting of $\Delta R$ is always of about $1.80 \pi$, independent of redshift $z$. For different cosmological parameters, $\theta$ also share a similar value, of about $1.77\pi$. Furthermore, in terms of the fitting for $\Delta R$, the magnitude of $C_2$ dominates over that of $C_1$, which is almost zero within uncertainty, as can be seen in Table 4. Therefore, $v_2 \equiv -\mathrm{sin}\theta \cdot m_{\nu} + \mathrm{cos}\theta \cdot \eta^2 \approx 0.6 m_{\nu} - 0.8 \eta^2$ should be the main contributor to the change in total matter power spectrum. The redshift dependences of $-C_m$, $C_\eta$, $C_1$ and $-C_2$ are plotted in Fig. 4d.

\begin{figure}
\subfigure[Fig. 5a]{
\includegraphics[width=7cm, height=6cm]{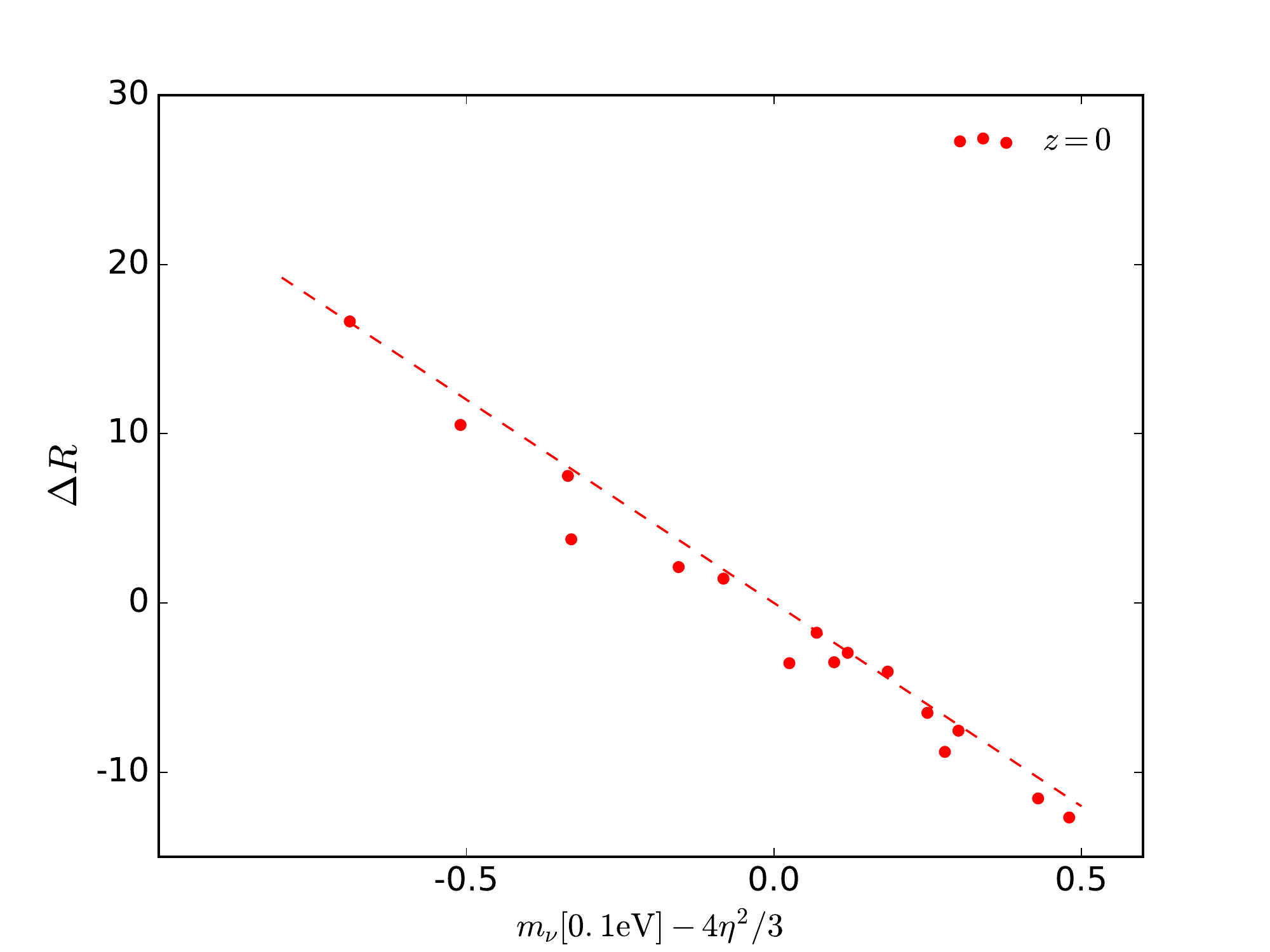}
}
\subfigure[Fig. 5b]{
\includegraphics[width=7cm, height=6cm]{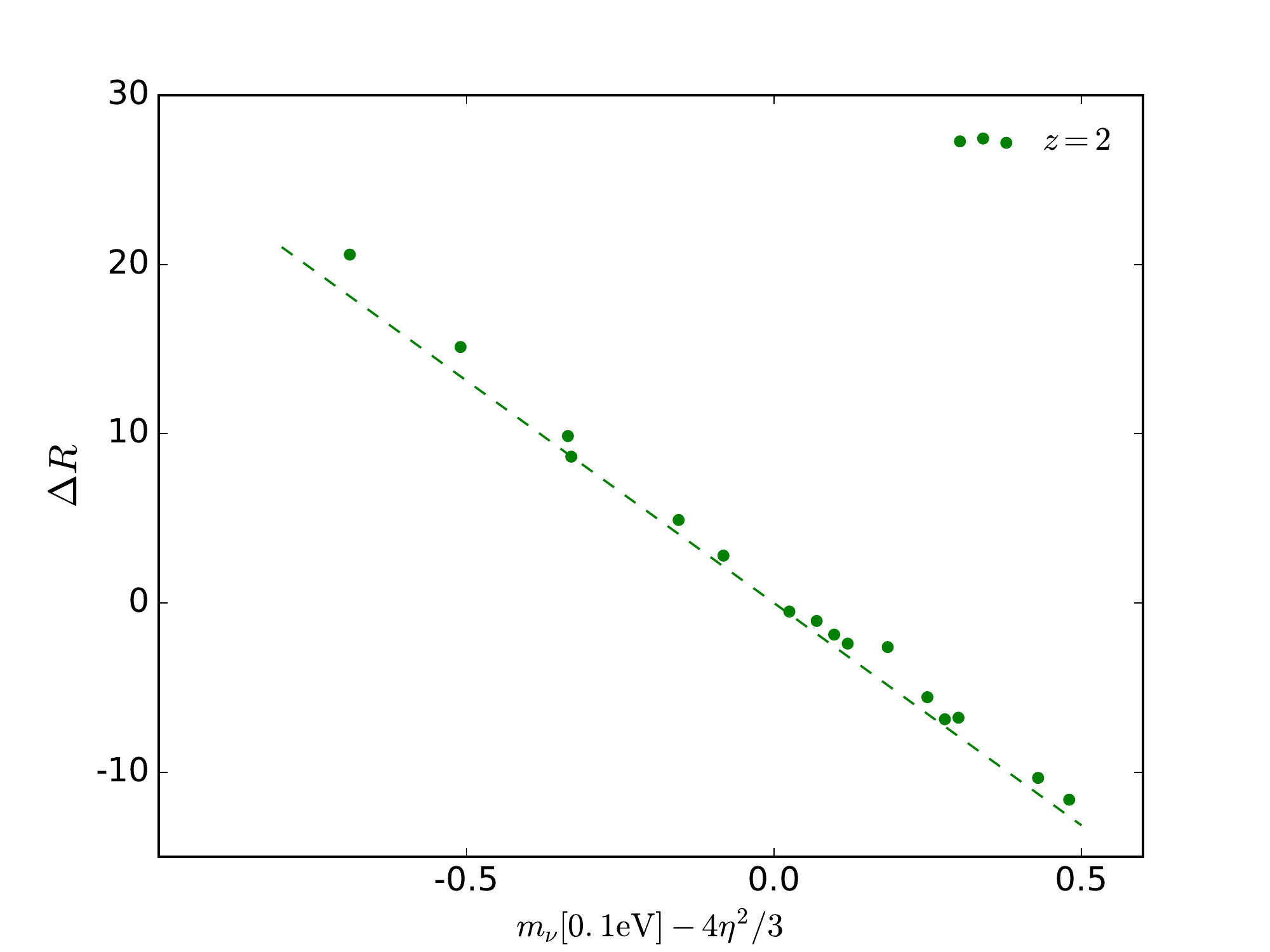}
}
\caption{\label{fig6}  a) At redshift $z=0$, represent the $\Delta R$ of C1-C16 as a function of $v_2 = m_\nu - \frac{4}{3} \eta ^2$ only. b) Same as a), but at redshift $z=2$, where the linear coefficient $C_1$ for another sub-dominant vector $v_1 = \frac{4}{3} m_\nu + \eta^2$ is closest to 0.}

\end{figure}

With our discovery that the change to matter power spectrum $\Delta R$ is mainly contributed by $v_2 = m_\nu - \frac{4}{3} \eta ^2$, we revisit Fig. 4b, but now plot $\Delta R$ as a function of $v_2$ only. In Fig. 5a we show the case of $z=0$, in comparison to Fig. 4b, while in Fig. 5b we show the case of $z=2$, where the linear coefficient $C_1$ for another vector $v_1 = \frac{4}{3} m_\nu + \eta^2$ is closest to 0. We can see that as expected, $v_2 = m_\nu - \frac{4}{3} \eta ^2$ is indeed the dominant contributor to the neutrino effects on the matter power spectrum.

\section{Summary and disscussion}

In this paper we have studied the impact of massive and degenerate cosmological neutrinos on the total matter power spectrum. Apart from the neutrino free-streaming effect and modified expansion history, we have also used the  \texttt{CosmoMC} code to refit the cosmological parameters from the Planck 2015 CMB data when finite neutrino mass and degeneracy are allowed, so that the study is self-consistent. Our simulations show that for a reasonable range of parameter values, the neutrino degeneracy parameter squared $\eta^2$ has a comparable effect on the matter power spectrum as that of the mass parameter $m_{\nu}$, but with opposite signs, and so there could be an issue of parameter degeneracy. Thus previous studies that estimate the neutrino mass purely from the free-streaming effect on LSS may not be accurate and should be re-evaluated. We provide a numerical fitting for the percentage deviation from $\Lambda$CDM of the averaged relative power spectrum $\Delta R = \bar{R}_{k[0.3, 1.0]} -1$ in the $k$-range $[0.3, 1.0]\ h\textrm{Mpc}^{-1}$, as a linear function against $m_{\nu}$ and $\eta^2$. The selection of this $k$-mode range can be customized for different observations. This two-variable regression shows good linearity. The redshift evolution of the corresponding coefficients $C_m$ and $C_{\eta}$ are also studied. We observe that although $C_m$ and $C_{\eta}$ are similar in magnitude in the late time, they differ at higher redshifts. We further investigate the covariance matrix of this numerical fitting and find that the off-diagonal term is comparable to the diagonal terms. Therefore, we propose to characterize the neutrino properties by a redshift independent parameter $m_{\nu} - \frac{4}{3}\eta^2$, which dominates the neutrino effects on the cosmological structure as well as the covariance. This also suggests that LSS alone may not be enough to break the parameter degeneracy between $m_{\nu}$ and $\eta$, and combined analyses with other cosmological probes such as CMB are needed.

We choose the grid-based method to include massive neutrinos in our simulation. It is not only more efficient, but can also be easily extended to neutrino models with a non-zero lepton asymmetry parameter $\eta$, which may not be compatible with the particle-based simulation schemes. This is because the non-zero $\eta$ effectively introduces a degeneracy pressure term apart from gravity, which is not a two-body force, and thus difficult to simulate in current N-body schemes. Our code is conceptually similar to that of \cite{ab2013}, only that we have modified \texttt{Gadget2} instead of \texttt{Gadget3}, which makes it much easier to be made public in the future. 

In this work, we investigate the effects of neutrinos on the matter power spectrum in the comoving units, e.g. $k[h \textrm{Mpc}^{-1}]$, which is the intrinsic unit in cosmological simulations, and thus the systematic error is minimized. However, as the fitting result of $h$ is dependent on $m_\nu$ and $\eta$, we may need to transform to physical units in future comparison with observations.

It should also be noted that as mentioned in \cite{barenboim_mass_eigenstate}, by the time of neutrino decoupling, the neutrino lepton asymmetry is already diagonal in mass eigenstates. Thus strictly speaking, cosmological neutrinos do not follow thermal distribution in mass eigenstates, but linear combination of thermal distributions in flavor eigenstates. Further studies into how reliable this assumption can be and how to improve it should be done. The authors of a recent study \cite{inman} comment on the grid-based simulation that the effect of the sound speed is ignored when the neutrinos are treated as fluid, which leads to a deficit in the neutrino power spectrum. This should not affect our results much, as the neutrino power spectrum only accounts for a tiny part in the total matter power spectrum. However, future studies may need to include this correction when high accuracy is required.

\appendix
\section{Linear evolution of the neutrino over-density}

The following derivation is mainly based on \cite{xiang}, with some reference to \cite{ab2013} as well.

The general continuity equation for a density field is given by 

\begin{equation}
\int \frac{\partial \rho}{\partial t} d^3r + \oint_V \rho \textbf{v} \cdot d\textbf{s} = \int \frac{\partial \rho}{\partial t} d^3r + \int \nabla \cdot (\rho \textbf{v}) d^3r = 0.
\end{equation}

So we have 

\begin{equation}
\frac{\partial \rho}{\partial t} + \frac{\partial (\rho \dot{r_i})}{\partial r_i} = 0.
\end{equation}

Considering also the momentum space, we substitute $\rho$ with $F_{\nu}$, the neutrino distribution function, defined by $\rho \propto \int F_{\nu} d^3p$, and Eq.(A.2) becomes

\begin{equation}
\frac{\partial F_{\nu}}{\partial t} + \frac{\partial (F_{\nu} \dot{r_i})}{\partial r_i} + \frac{\partial (F_{\nu} \dot{p_i})}{\partial p_i}= 0.
\end{equation}

Applying the following identity from the Hamiltonian equation 

\begin{equation}
\frac{\partial \dot{r_i}}{\partial r_i} + \frac{\partial \dot{p_i}}{\partial p_i} = \frac{\partial ^2 H}{\partial r_i \partial p_i} - \frac{\partial ^2 H}{\partial p_i \partial r_i} = 0,
\end{equation}

we obtain the Vlasov equation from Eq.(A.3): 

\begin{equation}
\frac{d F_{\nu}}{dt} = \frac{\partial F_{\nu}}{\partial t} + \frac{\partial F_{\nu}}{\partial r_i} \dot{r_i}+ \frac{\partial F_{\nu}}{\partial p_i} \dot{p_i} = 0.
\end{equation}

Here we use Newtonian gravity for the $\dot{p_i}$, because both the Vlasov equation and the \texttt{Gadget2} N-body simulation code are in the non-relativistic regime. So we have 

\begin{equation}
\dot{\textbf{p}} = m\nabla \phi = -m G\int \rho_{t} \frac{\textbf{r-r'}}{\textbf{|r-r'|}^3} d^3 r',
\end{equation}

where $\rho_{t}$ is the total energy density of all massive particles, including  cosmological neutrinos, baryons and CDM.

We can divide $F_{\nu}$ into an unperturbed Fermi-Dirac term $f^0_{\nu} (v)$ and a perturbation term which is position dependent, $f'_{\nu} (\textbf{r}, \textbf{v})$:

\begin{equation}
F_{\nu} = f^0_{\nu} (v) + f'_{\nu} (\textbf{r}, \textbf{v}).
\end{equation}

Then we put Eq.(A.6) and (A.7) into Eq.(A.5) and switch to the comoving coordinates:

\begin{equation}
\begin{split}
ds &= \frac{dt}{a^2(t)} \\
\textbf{x} &= \frac{\textbf{r}}{a(t)}  \\
\textbf{u} &\equiv \frac{d\textbf{x}}{ds} = a(t)\textbf{v} - Ha(t)\textbf{r}, 
\end{split}
\end{equation}

and obtain

\begin{equation}
\frac{\partial f'_{\nu}}{a^2\partial s} + \frac{\textbf{u} \cdot \partial f'_{\nu}}{a^2\partial \textbf{x}} - \ddot{a}a\textbf{x} \cdot \frac{\partial f^0_{\nu}}{\partial \textbf{u}} - Ga^2\frac{\partial f^0_{\nu}}{\partial \textbf{u}} \cdot \int \rho_{t}(s,\textbf{x'}) \frac{\textbf{x-x'}}{\textbf{|x-x'|}^3} d^3 x' = 0.
\end{equation}

We notice that there is an $\ddot{a}$ term in Eq.(A.9). Applying the Friedmann equation 

\begin{equation}
\frac{\ddot{a}}{a} = -\frac{4\pi G}{3} \bar{\rho} _{t},
\end{equation}

where $\bar{\rho} _{t}$ is the unperturbed total energy density, and 
\begin{equation}
\frac{4\pi \textbf{x}}{3} = \int \frac{\textbf{\textbf{x-x'}}}{\textbf{|x-x'|}^3} d^3 x', 
\end{equation}

then the last two terms in Eq.(A.9) can be combined. Finally we have 

\begin{equation}
 \frac{\partial f'_{\nu}}{\partial s} + \textbf{u} \cdot \frac{\partial f'_{\nu}}{\partial \textbf{x}} - Ga^4 \cdot \frac{\partial f ^0_{\nu}}{\partial \textbf{u}}\int \bar{\rho}_{t} \delta_{t}(s, \textbf{x'}) \frac{\textbf{\textbf{x-x'}}}{\textbf{|x-x'|}^3} d^3 x' = 0,
 \end{equation} 

where 
\begin{equation}
\bar{\rho}_{t} \delta_{t} \equiv \rho_{t} - \bar{\rho}_{t} = \bar{\rho}_{c+b} \delta_{c+b} + \bar{\rho}_{\nu} \delta_{\nu}.
\end{equation}

Applying Fourier transform to Eq.(A.12), we have

\begin{equation}
\frac{\partial \widetilde{f'_{\nu}} (s, \textbf{k}, \textbf{u})}{\partial s} + i\textbf{k}\cdot\textbf{u} \widetilde{f'_{\nu}} (s, \textbf{k}, \textbf{u}) - Ga^4 \frac{\partial f_{\nu}^0}{\partial \textbf{u}} \cdot \int [\bar{\rho}_{c+b}(s) \delta_{c+b}(s, \textbf{x'}) + \bar{\rho}_{\nu}(s) \delta_{\nu}(s, \textbf{x'})] d^3 x' \int e^{-i \textbf{k} \cdot \textbf{x}} \frac{\textbf{\textbf{x-x'}}}{\textbf{|x-x'|}^3}  d^3x = 0,
\end{equation}

where $\widetilde{f'_\nu}$ indicates the Fourier transformed perturbation of the neutrino distribution function.

Using the fact 

\begin{equation}
\int e^{-i \textbf{k} \cdot \textbf{x}} \frac{\textbf{x-x'}}{\textbf{|x-x'|}^3}  d^3x = e^{-i\textbf{k} \cdot \textbf{x'}} \int e^{-i\textbf{k} \cdot \textbf{y}}\frac{\textbf{y}}{\textbf{|y|}^3}  d^3y = -4\pi i \frac{\textbf{k}}{k^2}e^{-i\textbf{k} \cdot \textbf{x'}}
\end{equation}

to combine the first two terms in Eq.(A.14), and multiplying through by $e^{i\textbf{k}\cdot \textbf{u} s}$, we have 

\begin{equation}
\frac{\partial }{\partial s}[ \widetilde{f'_{\nu}}(s, \textbf{k}, \textbf{u}) e^{i\textbf{k} \cdot \textbf{u} s}] + e^{i\textbf{k}\cdot \textbf{u} s} 4\pi Ga^4 \frac{i\textbf{k}}{k^2}  \cdot \frac{\partial f_{\nu} ^0}{\partial \textbf{u}}  \int e^{-i\textbf{k} \cdot \textbf{x'}} [\bar{\rho}_{c+b}(s) \delta_{c+b}(\textbf{x'}) + \bar{\rho}_{\nu}(s) \delta_{\nu}(\textbf{x'})] d^3 x' = 0.
\end{equation}

Next we integrate Eq.(A.16) with respect to the time variable $s$, and we have 

\begin{equation}
\begin{split}
\widetilde{f'_{\nu}}(s, \textbf{k}, \textbf{u}) + \int _0 ^s e^{-i \textbf{k} \cdot \textbf{u} (s-s')} 4\pi Ga^4(s') \frac{i\textbf{k}}{k^2}  \cdot \frac{\partial f^0 _{\nu}}{\partial \textbf{u}} &[\bar{\rho}_{c+b} (s') \widetilde{\delta}_{c+b}(s', \textbf{k}) + \bar{\rho}_{\nu} (s') \widetilde{\delta_{\nu}}(s', \textbf{k})] ds' \\
&=  \widetilde{f'_{\nu}}(0, \textbf{k}, \textbf{u}) \cdot e^{-i\textbf{k} \textbf{u} s},
\end{split}
\end{equation}

where $\widetilde{\delta_{i}} (\textbf{k}) \equiv \int e^{-i\textbf{k}\cdot \textbf{x'}} \delta_i d^3 x'$ is the over-density field in k-space, and the term on the right hand side of Eq.(A.17) is the initial condition of this integration. 

Eq.(A.17) reveals the evolution of the perturbation of the neutrino distribution fuction, but we always need to turn to the over-density field, which is measurable. Integrating over $d^3 u$ on both sides, we have

\begin{equation}
\begin{split}
\widetilde{\rho}_{\nu}(s, \textbf{k})  +  \frac{i\textbf{k}}{k^2}  \cdot \int e^{-i \textbf{k} \cdot \textbf{u}(s-s')} \frac{\partial f^0 _{\nu}}{\partial \textbf{u}} d^3u & \int _0 ^s 4\pi Ga^4(s')  [\bar{\rho}_{c+b} (s') \widetilde{\delta}_{c+b}(s', \textbf{k}) + \bar{\rho}_{\nu} (s') \widetilde{\delta_{\nu}}(s', \textbf{k})] ds' \\
&= \int e^{-i\textbf{k} \cdot \textbf{u}s} \widetilde{f'_{\nu}}(0, \textbf{k}, \textbf{u}) d^3 u,
\end{split}
\end{equation}

where $\widetilde{\rho}_{\nu}(s, \textbf{k}) \equiv \widetilde{\delta}_{\nu}(s, \textbf{k}) \int f^0_{\nu} d^3u$. 

Integrating by parts, we have

\begin{equation}
 \int e^{-i \textbf{k} \cdot \textbf{u}(s-s')} \frac{\partial f^0 _{\nu}}{\partial \textbf{u}} d^3u = i \textbf{k} (s-s') \int e^{-i \textbf{k} \cdot \textbf{u} (s-s')} f_{\nu} ^0 d^3u.
\end{equation}

Also up to the first-order approximation, the perturbation to the initial distribution function can be treated as 

\begin{equation}
\widetilde{f'_{\nu}}(0, \textbf{k}, \textbf{u})=\int e^{-i\textbf{k}\cdot \textbf{x}} f'_{\nu}(0, \textbf{x}, \textbf{u})d^3x  \approx \int e^{-i\textbf{k}\cdot \textbf{x}} f^0_{\nu} (0, \textbf{u}) \delta_{\nu} (0, \textbf{x})d^3x =  f^0_{\nu} (0, \textbf{u})  \widetilde{\delta_{\nu}}(0, \textbf{k}).
\end{equation}

For convenience, we define 

\begin{equation}
\Phi (\textbf{q}) = \frac{\int f^0 _{\nu} e^{-i\textbf{q} \cdot \textbf{u}}d^3u} {\int f^0_{\nu} d^3u} .
\end{equation}

Then we put Eq.(A.19), Eq.(A.20) and Eq.(A.21) into Eq.(A.18), and we have the final equation of the neutrino over-density growth:

\begin{equation}
\widetilde{\delta_{\nu}}(s, \textbf{k}) = 4\pi G \int _0 ^s a^4(s')(s-s') \Phi [\textbf{k} (s-s')] [\bar{\rho}_{c+b} (s') \widetilde{\delta}_{c+b}(s', \textbf{k}) + \bar{\rho}_{\nu} (s') \widetilde{\delta_{\nu}}(s', \textbf{k})] ds' + \Phi (\textbf{k} s) \widetilde{\delta_{\nu}}(0, \textbf{k}).
\end{equation}

In Eq.(A.22), the second term on the right hand side is the linear part of the growth equation, so that the initial condition can be separated from a time dependent growth factor. The first term is a complicated self-involved integration, which reflects the interaction between the total gravitational potential and neutrino over-density field. Since $\widetilde{\delta _{\nu}} (s, \textbf{k})$ itself appears in the integral, we need to solve this integration equation iteratively. Eq.(A.22) can be rewritten as

\begin{equation}
\widetilde{\delta_{\nu}}(s, \textbf{k}) = F(s, \textbf{k}) + \int_0 ^s G(s, s', \textbf{k}) \widetilde{\delta_{\nu}}(s', \textbf{k}) ds'. 
\end{equation}

This equation is a Volterra equation of the second kind. The kernel 
$G(s, s', \textbf{k}) $ is of the same order as $10^{-3}a H_0^2  (s-s')\Phi[\textbf{k} (s-s')]$, where we can see $\Phi[\textbf{k} (s-s')] <1$ from Eq.(A.21), and for a small time step $H_0 \cdot s \ll 1$. Therefore if we use $F(s, \textbf{k})$ to be the initial seed, the iteration 

\begin{equation}
\begin{split}
\widetilde{\delta_{\nu}} ^{(n)}(s, \textbf{k}) = F(s, \textbf{k}) &+ \int_0 ^s G(s, s', \textbf{k}) \widetilde{\delta_{\nu}}^{(n-1)}(s', \textbf{k}) ds' =  F(s, \textbf{k}) + \int_0 ^s G(s, s', \textbf{k}) F(s', \textbf{k}) ds' \\
& + \iint _0^s G(s, s', \textbf{k}) G(s', s'', \textbf{k}) F(s'', k)ds' ds'' + ...
\end{split}
\end{equation}

would converge very fast. In our actual calculation, one iteration would be enough to achieve the accuracy of about $10^{-5}$.

Next let us focus on the evaluation of $\Phi (\textbf{q})$. In Eq.(A.21) the denominator can be evaluated by numerical integration. But for the numerator,

\begin{equation}
\int f^0 _{\nu} e^{-i\textbf{q} \cdot \textbf{u}}d^3u = 2\pi \int _0 ^\infty \int _0^{\pi} \frac{u^2[\cos(qu \cos \theta) - i\sin(qu\cos\theta)]\sin\theta}{e^{\frac{mu}{T} -\xi} +1} du d\theta + anti.,
\end{equation}

where $anti.$ means the contribution of anti-neutrinos. The sine part finally vanishes when integrating over $\theta$, and there remains 

\begin{equation}
\begin{split}
2\pi \int _0 ^\infty \int _0^{\pi} \frac{u^2 \cos(qu \cos\theta) \sin \theta}{e^{\frac{mu}{T} -\xi} +1} & du d\theta + anti. = 2\pi \int _0 ^\infty \frac{2u^2 \sin(qu) }{qu(e^{\frac{mu}{T} -\xi} +1)} du + anti. \\
&=4\pi \frac{T^2}{qm^2}[\int _0 ^\infty \frac{x \sin(Ax) }{(e^{x -\xi} +1)} dx + \int _0 ^\infty \frac{x \sin(Ax) }{(e^{x +\xi} +1)} dx]
\end{split},
\end{equation}

where $x = u\frac{m}{T}$ and $A = q\frac{T}{m}$. Some public numerical libraries such as $\texttt{gsl.qag}$ would fail to evaluate this integration, as there is a high frequency oscillation in the integral kernel when $x$ is large. Thus a series expansion is used for Eq.(A.26). For the anti-neutrino term, we simply use the geometric series:

\begin{equation}
\begin{split}
\frac{1}{e^{x +\xi} +1} = \frac{1}{e^{x +\xi}[1- (-e^{-(x +\xi)})]} = e^{-(x + \xi)} \sum _{n=0}^{\infty} (-1)^n e^{-n(x+\xi)} = \sum _{n=1} ^{\infty} (-1)^{n+1} e^{-n(x+\xi)}
\end{split}.
\end{equation}

Then after integrating over $x$, we have

\begin{equation}
 \int _0 ^\infty \frac{x \sin(Ax) }{(e^{x +\xi} +1)} dx = \sum _{n=1}^{\infty} (-1)^{n+1} e^{-n\xi} \frac{2nA}{(A^2 + n^2)^2}.
\end{equation}

But for the neutrino part, the integration kernel is a bit tricky to be expanded, because the convergence radius for $(-1)^{n+1} e^{-n(x-\xi)}$ does not cover the whole space. So we separate the integration into two parts

\begin{equation}
\int _0 ^\infty \frac{x \sin(Ax) }{(e^{x - \xi} +1)} dx = \int _0 ^\xi \frac{x \sin(Ax) }{(e^{x -\xi} +1)} dx + \int _0 ^\infty \frac{(y+\xi) \sin[A(y+\xi)]) }{(e^y +1)} dy. 
\end{equation}

The first term is evaluated numerically, where $x$ is small enough to avoid  the high oscillation problem. For the second part, we can again use the same series expansion. We define

\begin{equation}
\begin{split}
B_1(n) &\equiv \int _0 ^\infty e^{-ny} \cos(Ay) dy = \frac{n}{A^2 + n^2} \\
B_2(n) &\equiv \int _0 ^\infty e^{-ny} \sin(Ay) dy = \frac{A}{A^2 + n^2} \\
B_3(n) &\equiv \int _0 ^\infty y e^{-ny} \cos(Ay) dy = \frac{n^2-A^2}{(A^2 + n^2)^2} \\
B_4(n) &\equiv \int _0 ^\infty y e^{-ny} \sin(Ay) dy = \frac{2nA}{(A^2 + n^2)^2}
\end{split},
\end{equation}

and then 

\begin{equation}
\int _0 ^\infty \frac{(y+\xi) \sin[A(y+\xi)]) }{(e^y +1)} dy = \sum _1 ^{\infty} (-1)^{n+1} [\xi B_1(n) \sin(A\xi) + \xi B_2(n) \cos(A\xi) + B_3(n) \sin(A\xi) + B_4(n) \cos(A\xi)].
\end{equation}

Therefore, combining Eqs.(A.28)-(A.31) with Eq.(A.26), and dividing Eq.(A.26) with the normalization integration in Eq.(A.21), we finally have the expression for $\Phi (\textbf{q})$:

\begin{equation}
\Phi (\textbf{q}) = \frac{B_0 + \sum _{n=1}^{\infty}(-1)^{n+1} \{ \xi B_1(n) \sin(A\xi) + \xi B_2 (n) \cos(A\xi) + B_3 (n) \sin(A\xi) + B_4 (n) [\cos(A\xi) + e^{-n\xi}] \} }{A(\int _0 ^\infty \frac{x^2 }{e^{x-\xi} +1} dx + \int _0 ^\infty \frac{x^2 }{e^{x+\xi} +1} dx)},
\end{equation}

with $B_0 \equiv \int _0 ^\xi \frac{x \sin(Ax) }{(e^{x -\xi} +1)} dx$ and the denominator to be calculated by direct integration. 

\section{Error analysis of the fitting of $\Delta R (m_\nu, \eta^2)$}

There are four sources of errors in our final fitting of $\Delta R (m_\nu, \eta^2)$: a) the uncertainties of the refitting of cosmological parameters from CMB; b) the uncertainty in the measurement of the matter power spectrum; c) the uncertainty propagation from $P(k)$ to $\Delta R$; d) the uncertainty of fitting $\Delta R$ into a linear function of $m_\nu$ and $\eta^2$. Their corresponding treatments are elaborated as follows:

a) There are five cosmological parameters involved in our modified simulations, $\Omega_b$, $\Omega_c$, $H_0$, $A_s$ and $n_s$. First, we test how each affects the power spectrum using the halofit model in \texttt{CAMB}. It turns out that in the selected wavemode range $k \in [0.3, 1.0]\ h\textrm{Mpc}^{-1}$, increasing all of these parameters except for $\Omega_b$ would lead to an increase in the matter power spectrum. However, since in the N-body simulation, $\Omega_b$ and $\Omega_c$ are not distinguished when simulation particles are CDM only, we should consider their effects together. As a result, a large value of $\Omega_b + \Omega_c$ leads to a larger power spectrum. Therefore, for each simulation setup in Table 3, we run two additional data sets with the same $(m_{\nu}, \eta)$, but the cosmological parameters are chosen to be the $\pm 1\sigma$ values from our \texttt{CosmoMC} fitting. The power spectra measured from these simulations are denoted as $P(k)^+$ and $P(k)^-$ respectively.

b) The uncertainty of measurement of power spectrum is taken to be the statistical standard deviation of binning. As a conservative estimation, we choose the larger of $|P(k)^+ + \sigma_ {P(k)^+} -P(k)|$  and $|P(k)^- - \sigma_{ P(k)^-} -P(k)|$ to be the uncertainty of $P(k)$.

c) $\Delta R = \bar{R}_{[0.3, 1.0]} -1$ is the average of the deviation of the relative power spectrum $R(k) = P(m_\nu, \eta, k)/P(0, 0, k)$ from 1, and so we regard this ratio at each bin of $k$ as an independent measurement of $\Delta R$. Thus the propagation of uncertainty is given by

\begin{equation}
\sigma_{\Delta R} = \sqrt{ \frac{\sum [\sigma_ {R(k)}]^2 }{N^2}},
\end{equation}

where $N$ is the number of bins, and $\delta R(k)$ is given by the propagation 

\begin{equation}
\sigma_{ R(k)}  = \sqrt{\frac{1}{P^2(0, 0, k)} [\sigma_ {P(m_\nu, \eta, k)}]^2 + \frac{P^2(m_\nu, \eta, k )}{P^4(0, 0, k)} [\sigma_ {P(0, 0, k)}]^2}.
\end{equation}

d) Then the series of $\Delta R(m_\nu, \eta)$, together with their error bars, are fitted into Eq.(4.3) using \texttt{scipy.optimize.curve\_fit}. The corresponding covariance matrices are also given.

The uncertainties of fitting the changes on cosmological parameters into functions of $(m_\nu, \eta^2)$ as Eq.(4.1) are analyzed in a similar process, and the results are listed in Table 5.

\section{Refitting of cosmological parameters: correlation with $\eta$}

\begin{figure}
   \centering
   \includegraphics[width=\textwidth]{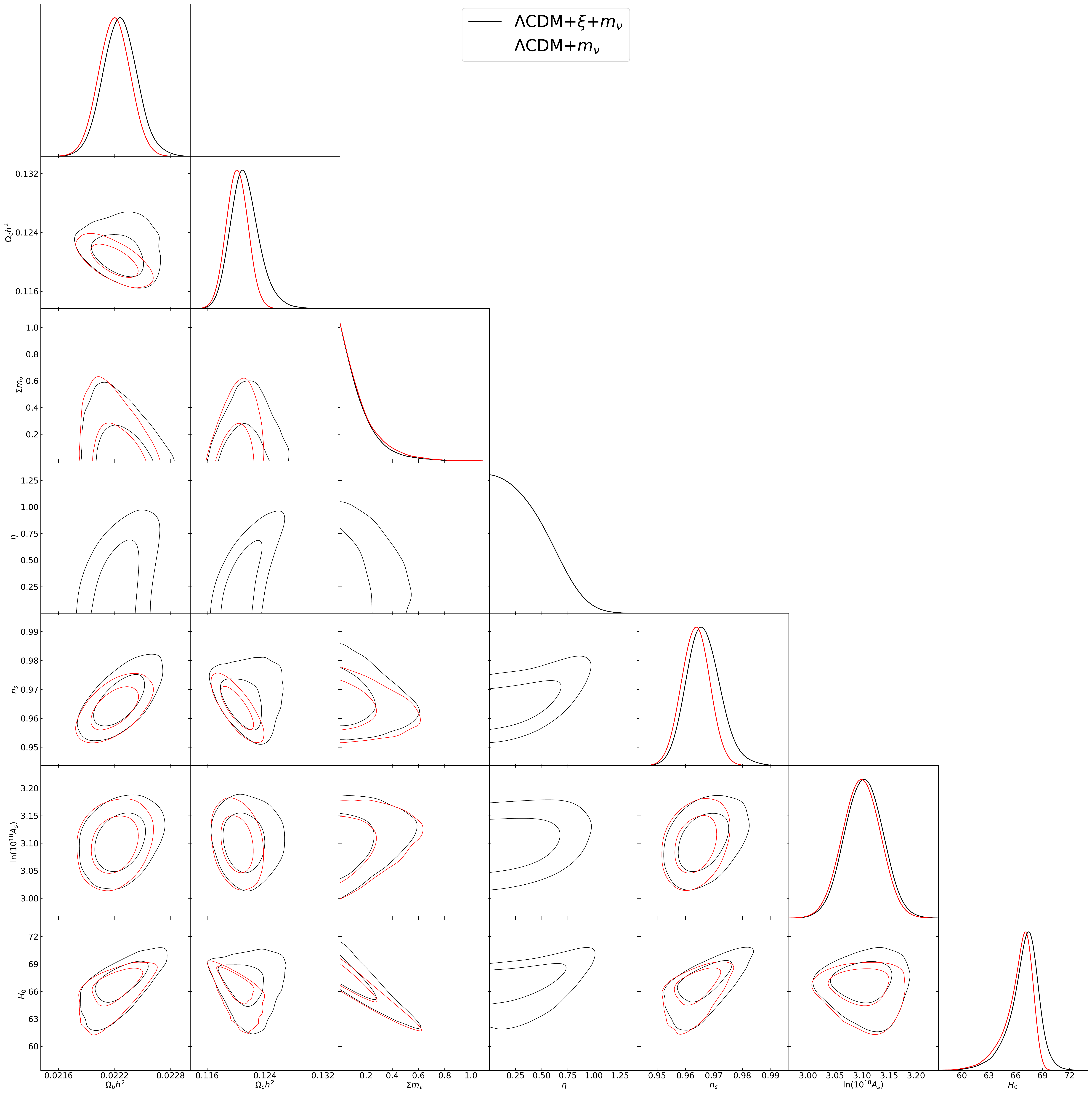}
   \caption{1D posterior pdfs and 2D contours (for 68\% CL and 95\%
CL) of selected cosmological parameters. The data sets
involved in this fitting are \texttt{lowTEB} and
\texttt{plikHM\_TTTEEE}, and the models are $\Lambda$CDM+$m_\nu$ (red
lines) and $\Lambda$CDM+$m_\nu+\eta$ (black lines).}
   \label{fig:tri_plot}
\end{figure}

As mentioned in the previous sections, $\eta$ increases the energy
density of neutrino and affects the expansion history of the universe.
Therefore we expect $\eta$ to be correlated with other cosmological
parameters, and the posterior distributions of all the cosmological
parameters would be affected.
We use the MCMC fitting method which is widely used by the CMB
community to find the posterior distribution of cosmological
parameters. The fitting is done by the \texttt{CosmoMC} code \cite{cosmomc} which
calculates the CMB anisotropy power spectrum by calling \texttt{CAMB},
a numerical solver of the Boltzmann equations.
We modified \texttt{CAMB} to calculate the power spectrum with a
non-zero $\eta$. \texttt{CosmoMC} is also modified to include the new
cosmological parameter $\eta$. We used the modified
\texttt{CosmoMC} together with the \textit{Planck} likelihood
code and the data sets \texttt{lowTEB} and \texttt{plikHM\_TTTEEE} for
the MCMC fitting. The data set \texttt{lowTEB} contains low multipole
$\ell$ temperature and LFI polarization data, while
\texttt{plikHM\_TTTEEE} contains high multipole $\ell$ temperature and
E mode polarization power spectra produced using cross half-mission
map. In addition to the standard cosmological parameters
$\{\Omega_\mathrm{c}h^2,~\Omega_\mathrm{b}h^2,~\theta,~\tau,~A_\mathrm{s},~n_\mathrm{s},m_\nu\}$,
we have added one extra parameter $\eta$. The posterior distributions
of the parameters are shown in Figure \ref{fig:tri_plot}.

\begin{figure}
   \centering
   \includegraphics[width=\textwidth]{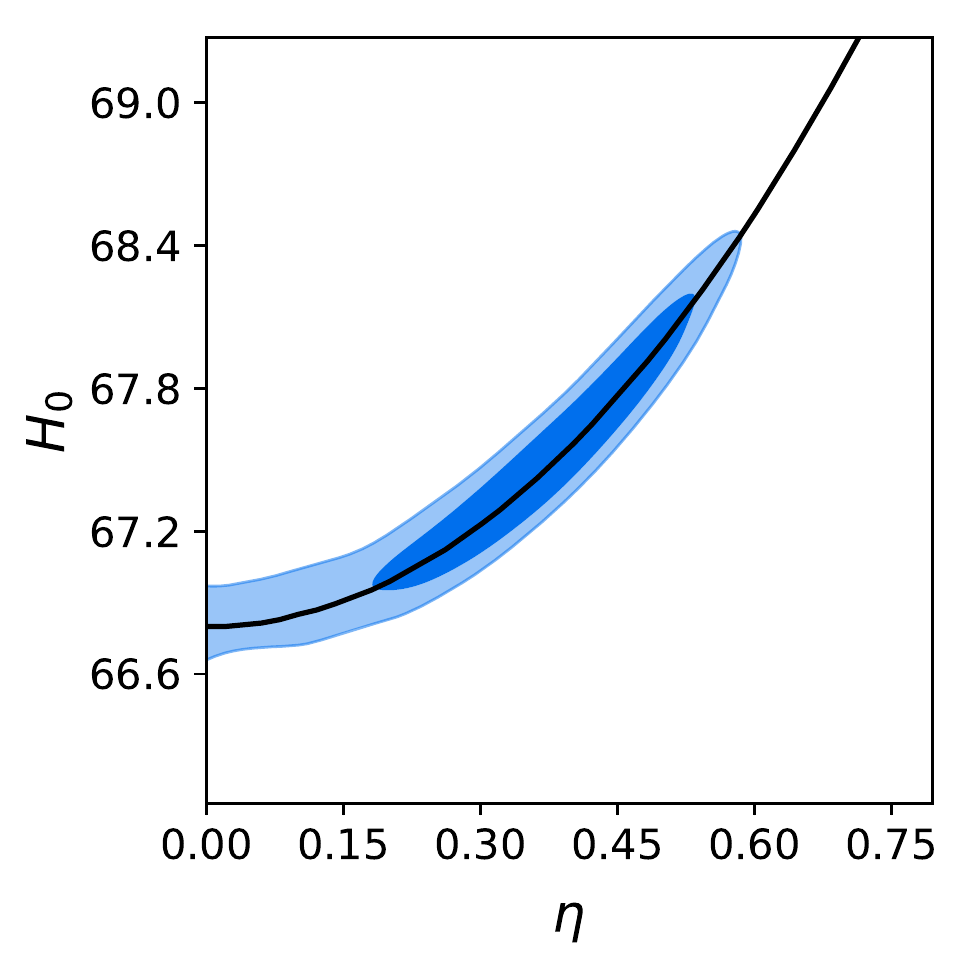}
   \caption{2D contours (for 68\% CL and 95\% CL) of $H_0$ versus
$\eta$. The data sets are the same as Figure \ref{fig:tri_plot}, and
all parameters except $H_0$ and $\eta$ are fixed. The black line is
the numerical constraint given by fixing $\theta_s$.}
   \label{fig:2D_plot}
\end{figure}

The results without adding $\eta$ are also plotted for comparison. As
expected, some of the parameters are correlated with $\eta$.
Therefore, the marginalized distributions of the parameters are also
affected. This also means that the mean values of the cosmological
parameters are different.
As an example, consider the correlation between $H_0$ and $\eta$. The
2D contours in Figure \ref{fig:tri_plot} show a clear positive
correlation between $H_0$ and $\eta$. This correlation mainly comes
from the tight constraint of the characteristic angular scale of the
CMB anisotropy $\theta_s$, defined as the ratio between $r_s$, the
comoving sound horizon at CMB decoupling and $D_A$, the comoving
distance of the last scattering surface. $\eta$ mainly affects the
early time expansion which determines $r_s$, while $H_0$ affects
the late time expansion which determines $D_A$. Since $\theta_s$
is tightly constrained, any change in $r_s$ must be compensated by a
corresponding change in $D_A$. As a result, the 2D contours of $H_0$
versus $\eta$ roughly follow a line given by the constraint
$\theta_s=\mathrm{const}$. To demonstrate this correlation, we
performed a MCMC fitting with all parameters except $\eta$ and $H_0$
fixed. The results are shown in Figure \ref{fig:2D_plot}.
More details of the MCMC fittings can be found in (Lau et al. in preparation).

\acknowledgments

We appeciate the inspiring discussions with Jiajun Zhang, Hantao Liu, Jianxiong Chen, Shihong Liao and Tom Broadhurst. Zeng would also like to thank Jiajun Zhang for his mentorship in learning about the N-body simulation. This work is supported partially by a CUHK VC Discretionary Fund. This work is partially supported by a grant from the Research Grant Council of the Hong Kong Special Administrative Region, China (Project No. 14301214). Simulations in this work were performed using the Central Research Cluster at CUHK.

% The bibliography will probably be heavily edited during typesetting.
% We'll parse it and, using the arxiv number or the journal data, will
% query inspire, trying to verify the data (this will probalby spot
% eventual typos) and retrive the document DOI and eventual errata.
% We however suggest to always provide author, title and journal data:
% in short all the informations that clearly identify a document.

\end{document}